# Spin Transport in Organic Semiconductors: A Brief Overview of the First Eight Years

Kazi M. Alam and Sandipan Pramanik[1]

Department of Electrical and Computer Engineering

University of Alberta, Edmonton, AB T6G 2V4, Canada

## Abstract

In this article we review the current state of the experimental research on spin polarized transport in organic semiconductors. These systems, which include small molecular weight compounds and polymers, are central in the rapidly maturing area of organic electronics. A great deal of effort has been invested in the last eight years toward understanding spin injection and transport in organics. These developments have opened up the possibility of realizing a new family of organic spintronic devices which will blend the chemical versatility of organic materials with spintronic functionalities.

---

[1] Corresponding author. Email: pramanik@ece.ualberta.ca

# Contents

## 1. Introduction

Processing and storing of binary data are central to information technology. These functions are often implemented by manipulating the charge degree of freedom of carriers (electrons and holes) in solid state systems. For example, realization of the classical binary logic bits ("0" and "1") requires two physically distinguishable states. In computer memories such states are realized by different amounts of charge stored on a capacitor[2] or by two distinct voltage levels at some circuit node[3]. Processing of such charge based logic bits is performed by circuits comprised of switching devices such as metal oxide semiconductor field effect transistors (MOSFETs). An emerging technology dubbed "spintronics"[4] [1, 2], on the other hand, aims to harness the spin degree of freedom of carriers in lieu of charge to realize these core information processing and storage functionalities [3-6]. This technology has already revolutionized the storage density of hard drives [7, 8] and can potentially realize universal memory [9] and low-power computation [3, 10].

In this article, our focus is on *spin polarized transport* in *organic semiconductors*. The first report on electrical spin injection and transport in organic semiconductor thin films appeared in 2002 [11, 12] and since then these materials have attracted significant interest [13-16]. Initial reports indicate that organic semiconductors have unique properties which can be fruitfully harnessed for spintronic information processing, data storage and novel opto-spintronic applications. Here we first briefly review the relevant concepts for spin transport in organic systems and then survey the main experimental results obtained during the first eight years of activity in this field with the aim of reconciling the available experimental data with theory.

A comprehensive review on organic spintronics [13], addressing both theoretical and experimental aspects, appeared in 2007 and covers the major results published up to that date. A more recent (2009) article [14] reviews the major experimental results. Ref.s [15, 16] mainly focus on more specialized areas such as organic nanowires and molecules [15] and work done at the University of Utah [16].

---
[2] e.g. dynamic random access memory or DRAM cells
[3] e.g. static random access memory or SRAM
[4] Acronym for "spin electronics", "spin based electronics" or "spin transport electronics"

This article is organized as follows: in the rest of this section, we outline traditional *inorganic-based* spintronic information processing and storage technologies [5, 7, 17] because initial applications of organic spintronics are likely to emerge from these areas. Section 2 lays out a discussion on spin injection and transport in the context of organics. Spin transport experiments in organic thin films and nanowires are discussed in sections 3 and 4 respectively. We will conclude in section 5 by outlining certain areas where organic spintronics can potentially carve a niche.

Note that there exist several organic magnetic field effects (OMFE) [18-20] such as field dependence of electroluminescence, photoluminescence, photocurrent and electrical-injection current. The origin of OMFE is not clear yet and existing theories often invoke interactions between spin polarized entities (polarons and excitons) and the hydrogen hyperfine field [21]. However, these phenomena are not explicitly related to spin injection and transport and therefore will not be covered in this article.

## *1.1 Spintronics in data storage.*

(a) *GMR read heads*:

The discovery of the giant magnetoresistance (GMR) effect in the late 1980s [22, 23] is arguably the first major milestone in the history of spintronics [7, 24] and has a significant impact on the storage industry. This phenomenon, observed in ferromagnetic/nonmagnetic metallic multilayers (ref. Figure 1a), offers an efficient and scalable method of sensing very weak magnetic fields. This has enabled significant reduction of bit size on hard discs without sacrificing the quality of signal detection, has led to an enormous increase in the areal recording density which currently exceeds 500 Gbit/inch$^2$ [7, 8], and is fast approaching unprecedented terabyte densities. This development has also resulted in hard drives of smaller form factors which are now ubiquitous in virtually all mobile appliances such as ultra-light netbooks, portable multimedia players, digital cameras and camcorders.

(b) *Magnetic random access memory*:

Spintronic memory (magnetic random access memory or MRAM) chips[5] have also started to trickle into the memory market [25] and are emerging as a strong contender in the race for "universal memory" [9]. The primitive memory cell consists of a tunnel barrier of aluminum oxide [26, 27] or magnesium oxide [28, 29] separating two ferromagnetic electrodes (also known as the "magnetic tunnel junction" or MTJ). The resistance of the cell is determined by the relative magnetization orientations of the electrodes and can be in either of two well-separated non-volatile states ("high-resistance" or "low-resistance"). This device can therefore be utilized to store binary data. This method of storage is nonvolatile since the resistance depends on the relative magnetization orientations of the contacts which remain unaffected when the power is switched off. In this respect, MRAMs are similar to flash memories but offer superior read-write speed and endurance. A schematic description of MTJ is shown in Figure 1(b).

Apart from memory cells, MTJs are also finding application as read heads in hard drives [30]. The binary nature of the resistance states can also be exploited to develop logic gates [31].

## *1.2 Spintronics for information processing.*

The application of spintronics in the realm of information processing is a relatively new endeavor and is motivated by the belief that spintronics may offer a more power-efficient route compared to the traditional transistor based paradigm [32]. Since the inception of silicon-based integrated circuits ("chip") over four decades ago, the integration density of transistors on a chip has roughly doubled in every 18 months with concomitant increase in clock speed and computational prowess. This trend, commonly referred to as Moore's law, has been sustained by the combined effect of classical Dennard scaling [33] of transistors and gradual introduction of non-traditional materials (e.g. high $\kappa$ dielectric, metal gate) in the fabrication process [34]. If this trend continues, the packing density will be $\sim 10^{13}$ transistors per cm$^2$ in the year of 2025 [3]. This implies that the gate length of these transistors should be $\sim 1$ nm. Field effect devices with

---

[5] e.g. MR2A16A from Freescale Semiconductor and MR2A08AYS35 from Everspin Technologies

sub-10 nm gate length have already been reported [35, 36] and further improvements can be expected based on process and materials innovation [37]. The fundamental problem, however, is the issue of power dissipation [4, 38]. Standard heat sinking technologies such as convective [39] or thermoelectric cooling [40] are capable to remove $\sim 1 \, kW/cm^2$. If the projected clock speed is $\sim 10$ GHz, then to avoid chip meltdown the energy dissipation per bit flip must be restricted to 0.01 attoJoules. This is a major challenge for MOSFETs which currently dissipate $\sim 1$ fJ per bit flip [3]. If the technology is unable to reduce the dissipation associated with each bit flip, then the power dissipation constraint will impose limitation on the maximum number of transistors that can be packed on a chip and the maximum clock frequency [41]. This impasse (along with other potential technical show-stoppers) constitutes the so-called "Red Brick Wall" [42] on the International Technology Roadmap for Semiconductors (ITRS) and calls for ingenious designs to reduce power dissipation. This is where spintronics is expected to offer a bailout.

Spintronic information processing can be achieved in three conceptually different ways. These are
(*a*) classical information processing using a monolithic [43] (or "all-spin") approach [4, 44-46],
(*b*) classical information processing using a hybrid [43] (or "charge *augmented with* spin") approach [47-53] and
(*c*) quantum information processing using spin qubits [54-59].

In the monolithic approach [10, 43], charge does not play any direct role in information coding and processing. These tasks are performed by spins. The single spin logic (SSL) paradigm [3, 43, 44] is an example of this monolithic approach and utilizes the fact that in the presence of an external magnetic field, an electron spin can either be parallel ("upspin") or antiparallel ("downspin") to the magnetic field. This bistable nature of spin is exploited to encode classical logic bits "0" and "1". Processing of this information is performed by applying a local magnetic field and controlling spin-spin interactions (Figure 1c). Since this paradigm employs stationary charges and does not involve any physical movement of charge carriers (and hence no current), it eliminates the $I^2R$ type dissipation that accompanies every MOSFET with channel resistance *R* carrying a current of *I* during switching. Still a finite amount of energy is dissipated during the bit (spin) flip, but this is significantly smaller compared to the MOSFETs [3, 43, 60]. Similar proposals have been made which encode logic bits in the magnetizations of nanomagnets[6] i.e. binary data is now encoded by a collection of spins instead of a single spin [46, 61]. Here spin manipulation is performed via spin torque effect [46] or local magnetic fields [61]. By suitable design, power dissipation associated with a bit (magnetization) flip can be reduced to only few $k_BT$ [46].

In the so-called "hybrid" approach [43] the focus is to realize transistor switches in which the device current is turned on and off by controlling the spin orientations of the charge carriers (Figure 1d)[7]. The logic operation is otherwise same as the classical charge based schemes. It has however been argued that spin transistors are unlikely to offer any significant advantage over the conventional MOSFETs as far as the power dissipation is concerned [3, 60]. Ref. [3] provides an extensive review on classical information processing using spins.

Finally, the third approach proposes to use spin as a quantum bit (qubit) instead of a classical bit as in the single spin logic approach. A quantum computer manipulates qubits and offers immense computational prowess and zero energy dissipation [62].

*1.3 Organic semiconductor spintronics.*

Most of the studies in spintronic data storage and information processing described above have been performed on metallic systems (GMR, metallic spin valves) [7], tunneling insulators (MTJ) [7] and inorganic semiconductors (e.g. silicon [63, 64], gallium arsenide [65], indium arsenide [53]). Organic semiconductors are relatively new entrants in the domain of spintronics. These materials are mostly hydrocarbons with a $\pi$ conjugated structure with alternating single and double bonds. The $p_z$ orbitals of the $sp^2$ hybridized carbon atoms overlap to form a delocalized $\pi$ electron cloud. The electrons in this region do not belong to a single atom or bond, but belong to the entire conjugated molecule or polymer chain. Transport occurs via hopping between the neighboring states and this is responsible for the semiconducting properties exhibited by these materials. It is also possible to increase the conductivity of organics by doping [66].

---

[6] This is how data is encoded in hard disc drives.
[7] The traditional charge-based metal oxide semiconductor field effect transistor (MOSFET) performs this operation by an electrostatic "field effect".

Organic semiconductors can be classified in two broad categories: (*a*) low molecular weight compounds and (*b*) long chain polymers. These materials have already found a niche in the silicon dominated electronics market, their attractive features being inherent structural flexibility, availability of low-cost bulk processing techniques, integrability with inorganic materials and possibility of chemical modification of the molecular structure to obtain tailor-made optical and electrical properties. These materials have been used extensively to realize devices such as organic light emitting diodes [67, 68], photovoltaic cells [69-71], field effect transistors [72] and even flash memories [73]. Today "*organic electronics*" is an immensely active research field [68, 74-76] with novel commercial applications including portable "roll-up" large area displays and lightings, smart textiles, organic smart cards, labels, tags, transparent electronic circuits, solar cells and batteries.

From a spintronic perspective, organic semiconductors exhibit very weak spin-orbit interaction which results in long spin lifetimes in such systems [13, 77, 78]. Long spin lifetime is the linchpin for both single spin logic and spin based quantum computing described above. These materials can also be used as tunnel barriers in MTJ cells leading to flexible MRAMs. The electroluminescence intensity from an organic diode may be controlled by spin polarized injection of electrons and holes [79, 80]. Thus organic semiconductors have the potential to emerge as an ideal platform for spin based computing, storage and opto-spintronic applications [13-15]. We will discuss more about these applications in section 5.

## 2. Basic elements of spin transport and implications for organics

What follows here is a short overview of the basic components of spin transport, namely spin injection, relaxation and detection. While extensively studied in the context of technologically important inorganic semiconductors [17], these phenomena are still relatively poorly understood in organics. Indeed there exist reports which questioned the feasibility of spin injection in organics [81]. Fortunately this issue is now resolved and it is generally accepted that spin injection is indeed possible in organics [82]. However, the dominant spin relaxation mechanism in organics is still a topic of much discussion [21, 83-85].

*2.1 Spin injection.*

The basic idea of spin injection in a paramagnetic semiconductor is as follows. In a paramagnet, at equilibrium, the spin magnetic moments[8] of different charge carriers (electrons in the conduction band and holes in the valence band) point along random directions in space. Therefore the ensemble average spin moment at some arbitrary location inside the device at any given time is zero. In other words, the charge carriers are unpolarized. Under an applied electrical bias these unpolarized ensemble of charge carriers flow from one point to another and result in charge current. For "spin transport", however, the basic principle is to create a (non-equilibrium) population of carriers inside the device such that the net spin moment is non-zero and to manipulate this spin polarization via a gate electrode [47] or some local magnetic field for device operation. Creation of such a non-equilibrium situation is generally termed as "spin injection". There exist optical, transport and resonance methods to create a non-equilibrium population of spins. The transport method is more suitable for device applications and in this section we will only focus on *electrical injection* of spins. A detailed description of other methods of spin injection is available in [17, 86]. In case of organics, due to their inherently weak spin-orbit coupling, optical method is not a viable route for spin injection.

In an all-electrical spin transport experiment, spin injection is achieved by using ferromagnetic (transition metals, half-metals or dilute magnetic semiconductors) contacts, commonly described as "spin injectors". The magnetization of the ferromagnet, $M \propto n_\uparrow - n_\downarrow$ where $n_\uparrow(n_\downarrow)$ are the majority (minority) spin concentrations obtained by integration over the filled states. The majority (minority) spins have a magnetic *moment* parallel (antiparallel) to the magnetization. The carriers at the Fermi level are also spin polarized (in ideal half-metals, they are 100% spin polarized); so when an electric current flows from a ferromagnet to a paramagnet, this current is expected to be spin-polarized as well. This is the basis of *electrical spin injection* and is illustrated in Figure 2a. In this context it is useful to define two quantities (*a*) density of states (DOS) spin polarization $P_{DOS}$ of the injector ferromagnet and (*b*) spin polarization $\eta$ of the injected current (also known as the spin injection efficiency):

$$P_{DOS} = \left|\frac{\rho_\uparrow - \rho_\downarrow}{\rho_\uparrow + \rho_\downarrow}\right| \tag{1}$$

$$\eta = \frac{J_\uparrow - J_\downarrow}{J_\uparrow + J_\downarrow} \tag{2}$$

---

[8] The spin magnetic moment $\vec{\mu_s} = -g\mu_B \vec{s}/\hbar$ where $|\vec{s}| = \hbar/2$ is the spin angular momentum, $\mu_B$ is Bohr magneton and the g factor ($g$) is ~2 for free electrons and ferromagnetic metals such as Fe, Co and Ni.

In equation (1), $\rho_\uparrow(\rho_\downarrow)$ represents the density of states *at the Fermi level* of carriers with spin magnetic moments parallel (antiparallel) to the magnetization of the ferromagnet. Note that $P_{DOS}$ and $M$ do not necessarily have the same magnitude or even the same sign [87]. For Co and Ni, minority spin contribution is dominant at the Fermi level, making $P_{DOS}$ negative [88].

The two spin species may also have different mobilities and therefore *conductivity polarization* of the ferromagnet, defined as $P_{\sigma F} = (\sigma_\uparrow - \sigma_\downarrow)/(\sigma_\uparrow + \sigma_\downarrow)$, is not in general equal to $P_{DOS}$ and they may even have different signs. For Co and Ni, $P_{\sigma F}$ is positive. For Fe, $P_{DOS} > 0$ but $P_{\sigma F} < 0$.

In equation (2), $J_\uparrow(J_\downarrow)$ represents the current carried by the majority (minority) spin species in the paramagnet, immediately after spin injection. If the paramagnet is just a tunnel barrier then the spin polarization of the tunneling current (i.e. $\eta$) can be determined by Meservey-Tedrow technique [89]. This quantity is however not necessarily equal to $P_{DOS}$ since tunnel current not only depends on the DOS but also on the tunneling probability which may differ for different electronic states [88]. For example, ferromagnetic metals such as Co and Ni exhibit positive spin polarization of the tunnel current ($\eta > 0$, for alumina tunnel barrier), implying dominance of the majority spins in the tunnel current [88, 89]. This is in contradiction with their bulk band structures which show dominant contribution of the minority spins at the Fermi level, implying negative $P_{DOS}$. The barrier plays a crucial role in determining the tunnel probability and hence the sign of the spin polarization of the tunnel current. For example, ref. [90] showed that Co exhibits a negative spin polarization of tunneling electrons for SrTiO$_3$ barrier but a positive spin polarization for alumina barriers. In the organic realm, Co, Fe and permalloy ($Ni_{80}Fe_{20}$) exhibit a positive spin polarization of tunneling electrons for purely organic ($Alq_3$) or hybrid ($Al_2O_3/Alq_3$) barriers [91].

For spin injection from a ferromagnet to a paramagnet via an interfacial barrier, the spin injection efficiency ($\eta$) can be expressed as a weighted average of the conductivity polarizations of the bulk ferromagnet ($P_{\sigma F}$), the interface ($P_{\sigma i}$) and the bulk paramagnet ($P_{\sigma N}$) [17]. The last quantity ($P_{\sigma N}$) is zero by definition and the weights are proportional to the corresponding resistances:

$$\eta = \frac{r_F P_{\sigma F} + r_i P_{\sigma i}}{r_F + r_i + r_N} \tag{3}$$

where $r_{F,i,N}$ represent the effective resistance of the bulk ferromagnet, the interface and the bulk paramagnet respectively. In this equation $P_{\sigma i}$ is the conductivity polarization of the interface which depends on the nature of charge injection from the ferromagnet. The quantity $\eta$ indicates how efficiently spins are injected into the semiconductor from the ferromagnet, and hence the name "spin injection efficiency".

Some general observations can be made from the above equation. For an ohmic contact between a ferromagnetic and paramagnetic metal, $r_i = 0$, and $r_F \approx r_N$. Thus $\eta \approx P_{\sigma F}$, implying significant spin injection, as determined by the conductivity polarization of the ferromagnet. Spin injection has indeed been observed in all-metal structures [92-94]. However if the paramagnet is a semiconductor, then $r_F \ll r_N$ and if the contact is ohmic, $r_i = 0$. In this case $\eta \ll P_{\sigma F}$, implying poor spin injection. This is the well-known *conductivity mismatch problem* [95] that prohibits efficient spin injection from a ferromagnet to a semiconductor via an ohmic contact. However the effects of conductivity mismatch can be mitigated if dilute magnetic semiconductors or half-metallic ferromagnets are used as spin injector (instead of ferromagnetic metals). For these materials $r_F \approx r_N$, implying reasonably good spin injection ($\eta \approx P_{\sigma F}$). Alternatively, if there is a tunnel (Schottky) barrier at the metallic ferromagnet/semiconductor interface, then $r_i \gg r_F, r_N$, which will again result in efficient spin injection $\eta \approx P_{\sigma i}$ [96]. For a tunnel barrier, $P_{\sigma i} \neq 0$ since the spin-up and spin-down electrons at the Fermi level of the injector ferromagnet have different wavefunctions and hence different transmission probabilities. The tunnel barrier thus offers different conductivities to different spins. However, if carrier injection takes place via thermionic emission over the barrier then $P_{\sigma i} \approx 0$ since thermionic emission is essentially spin-independent. A recent theoretical study on organics [97] corroborates these observations. Indeed, we will see later in this article that for organic spin valves, due to the high resistivity of organic semiconductors, half metallic contacts and tunnel barriers are commonly employed to enhance spin injection.

*2.2 Spin relaxation.*

After injection, the spin polarized carriers travel through the (thick) paramagnetic semiconductor under the influence of a transport-driving electric field. In case of organics, the efficient spin injection occurs via an interfacial tunnel barrier as discussed above. The transport occurs either via the HOMO/LUMO levels or by hopping among the defect states within the HOMO-LUMO gap or by a combination of both. During their transit, different spins interact with their environments differently (spin-orbit, hyperfine and carrier-carrier interactions) and their original orientations get changed by various amounts. Thus the *magnitude (P) of the ensemble spin polarization* decreases with time ($t$) as well as with distance ($x$) measured from the injection point (ref. Figure 2b). This gradual loss in *magnitude* of the injected spin polarization is termed as *spin relaxation*. Assuming an exponential decay, spin relaxation/diffusion length $L_s$ (or spin relaxation/diffusion time $\tau_s$) is defined as the distance (duration) over which the spin polarization reduces to $1/e$ times its initial value. For $L \gg L_s$ (or $t \gg \tau_s$), $P \to 0$, implying complete loss of spin polarization. In other words, spin relaxation phenomenon in the paramagnet tends to bring the non-equilibrium spin polarization back to equilibrium unpolarized condition. In spintronics, since the underlying philosophy is to exploit the non-zero spin polarization, one always attempts to suppress spin relaxation *i.e.* enhance spin relaxation length and time in the paramagnetic semiconductor. Note that if the spins of all carriers change in unison, then the magnitude of the ensemble spin polarization will remain the same but its orientation will change. In this case spin relaxation is suppressed. The exponential decay of the injected spin polarization with distance has been demonstrated in [98].

There are several mechanisms in solid that are responsible for spin relaxation [99]. For example,

in case of organic/inorganic semiconductors and metals, the most dominant mechanisms are (*a*) Elliott-Yafet, (*b*) D'yakonov-Perel', (*c*) Bir-Aronov-Pikus and (*d*) hyperfine interaction with nuclei [17, 99]. Among these, the first two mechanisms accrue from spin-orbit interaction. The third one originates from exchange coupling between electron and hole spins, and the last is due to interaction between carrier spins and nuclear spins. It is not clear which of these mechanisms plays the most dominant role in organic semiconductors. These mechanisms are briefly described below.

(*a*) *Elliott-Yafet mechanism*

In the presence of spin-orbit coupling, Bloch states of a real crystal are not spin eigenstates ($|\uparrow\rangle$ or $|\downarrow\rangle$), but an admixture of both:

$$u_k(r) = a_k(r)|\uparrow> + b_k(r)|\downarrow> \qquad (4)$$

Therefore these states are not pure spin states with a fixed spin quantization axis, but are either pseudo-spin-up or pseudo-spin-down. The degree of admixture (i.e. the quantities $a_k, b_k$) is a function of the electronic wavevector $k$. As a result, when a momentum relaxing scattering event causes a transition between two states with different wavevectors, it will also reorient the spin and contribute to spin relaxation. This is the basis of the Elliott-Yafet mode of spin relaxation (Figure 3).

It is important to note that in case of Elliott-Yafet mechanism, the mere presence of spin-orbit interaction in the system does not cause spin relaxation. Only if the carriers are scattered during transport, spin relaxation takes place. Higher the momentum scattering rate, higher is the spin scattering rate [17].

This observation can be extended even in the case of hopping transport in disordered (non-crystalline) solids such as organics, where there is no band structure and no Bloch states as such, but momentum relaxation still causes concomitant spin relaxation. Even though spin-orbit interaction is weak in organics, momentum scattering rate is very high which can potentially make Elliott-Yafet a dominant contributor in spin relaxation.

(*b*) *D'yakonov-Perel' mechanism*

The D'yakonov-Perel' mechanism of spin relaxation is dominant in systems which lack inversion symmetry. Examples of such (inorganic) solids are III-V semiconductors (e.g. GaAs) or II-VI semiconductors (e.g. ZnSe) where inversion symmetry is broken by the presence of two distinct atoms in the Bravais lattice. Such kind of asymmetry is known as *bulk inversion asymmetry*. In a disordered organic semiconductor bulk inversion symmetry is absent. Inversion symmetry can also be broken by an external or built-in electric field which makes the conduction band energy profile inversion asymmetric along the direction of the electric field. Such asymmetry is called *structural inversion asymmetry*. In a disordered organic semiconductor, structural inversion asymmetry typically arises from microscopic electric fields due to charged impurities and surface states (e.g. dangling molecular bonds).

Both types of asymmetries result in effective electrostatic potential gradients (or electric fields) that a charge carrier experiences. In the rest frame of a moving carrier, this electric field Lorentz transforms to an effective magnetic field $B_{eff}$ *whose strength depends on the carrier's velocity $v$*. A carrier's spin in an inversion asymmetric solid undergoes continuous Larmor precession about $B_{eff}$ between two consecutive scattering events. Since the magnitude of $B_{eff}$ is proportional to the magnitude of carrier velocity $v$, it is different for different carriers. Thus, collisions randomize $B_{eff}$ and therefore the orientations of the precessing spins. As a result, the ensemble averaged spin polarization decays with time. This is the D'yakonov-Perel' mode of spin relaxation (Figure 3).

Now, if a carrier experiences frequent momentum relaxing scattering (*i.e.* small mobility and small momentum relaxation time $\tau_p$) then $v$ is small implying $|B_{eff}|$ is small and the spin precession frequency (which is proportional to $B_{eff}$) is also small. As a result, the D'yakonov-Perel' process is less effective in low mobility samples than in high mobility samples. It therefore stands to reason that the spin relaxation rate due to D'yakonov-Perel' process will be inversely proportional to the momentum scattering rate [17].

Note that the Elliott-Yafet and the D'yakonov-Perel' mechanisms can be distinguished from each other by the opposite dependences of their spin relaxation rates on mobility. In the former mechanism, the spin relaxation rate is inversely proportional to the mobility and in the latter mechanism, it is directly proportional. Indeed this argument has been used to show that D'yakonov-Perel' mode is not dominant in organic semiconductor $Alq_3$ [100]. We will describe this experiment in section 3.2.

(*c*) *Bir-Aronov-Pikus mechanism*

This mechanism of spin relaxation is dominant in bipolar semiconductors. The exchange interaction between electrons and holes is described by the Hamiltonian $H = A\, S.J\, \delta(r)$ where $A$ is proportional to the exchange integral between the conduction and valence states, $J$ is the angular momentum operator for holes, $S$ is the electron spin operator and $r$ is the relative position of the electron and the hole. Now, if the hole spin flips (owing to strong spin-orbit interaction in the valence band), then electron-hole coupling will make the electron spin flip as well, resulting in spin-relaxation of electrons. A more detailed description is available in reference [17].

In case of unipolar transport, *i.e.* when current is carried by either electrons or holes but not both simultaneously, this mode of spin relaxation is ineffective. But for spin based organic light emitting diodes (section 5) this mode may play a dominant role.

(*d*) *Hyperfine interaction*

Hyperfine interaction is the magnetic interaction between the spin magnetic moments of the electrons and the nuclei. This is the dominant spin relaxation mechanism for quasi-static carriers, i.e. when carriers are strongly localized in space and have no resultant momentum. In that case, they are virtually immune to Elliott-Yafet or D'yakonov-Perel' relaxations since these, as discussed above, require carrier motion. We can view the hyperfine mechanism as caused by an effective magnetic field ($B_N$), created by an ensemble of nuclear spins, which interacts with electron spins and results in dephasing (Figure 3). The hyperfine magnetic field $B_N$ depends weakly on the transport driving electric field and temperature.

Most inorganic semiconductors have isotopes that carry non-zero nuclear spins. *All* naturally occurring isotopes of $Ga$, $In$, $Al$, $Sb$ and $As$ have nuclear spins with substantial magnetic moments and hence nuclear hyperfine interaction is particularly dominant in technologically popular III-V semiconductors e.g. $(Al)GaAs$, $In(Al)As$ etc. On the other hand, materials like $Cd$, $Zn$, $S$, $Se$ and $Te$ have nuclear spin carrying isotopes with less natural abundance and hence in II-VI semiconductor materials nuclear hyperfine interaction is weaker compared to their III-V counterparts. Due to the same reason nuclear hyperfine interaction is weak in case of elemental semiconductors e.g. $Si$ and $Ge$. Hyperfine interaction can be completely avoided in case of II-VI semiconductors, $Si$ and $Ge$ if we use isotopically purified material containing a greatly reduced amount of nuclear spin carrying isotopes. But due to the high cost of isotopic purification, this line of investigation is rarely pursued.

In case of organic semiconductors, two major constituents are carbon and hydrogen. Carbon atoms do not contribute significantly to the hyperfine interaction since the most abundant (98.89%) isotope of carbon, $^{12}$C, has zero nuclear spin. Hyperfine interaction in organics mainly originates from the hydrogen atoms ($^{1}$H, abundance > 99.98%, nuclear spin $I = {}^{1}/{}_{2}$) with negligible contributions from $^{13}$C (natural abundance 1.109%) and other minor constituents. Interestingly, deuterium ($^{2}$H, natural abundance 0.015%) has much weaker hyperfine coupling strength (i.e. weaker $B_N$) and deuterated organics are expected to have substantially weaker hyperfine interaction. Such chemical modifications have been employed recently to investigate the role of hyperfine interaction in spin relaxation for various organics [21, 85]. Ref. [83] has reported a theory of hyperfine interaction mediated spin relaxation for organic semiconductors. This model predicts weak dependence of spin relaxation length $L_s$ on temperature and a rapid increase of $L_s$ with bias. These features can be used to identify hyperfine interaction, if it plays a dominant role in any organic [84, 101].

## *2.3 Spin relaxation in organics: General considerations.*

In most organics, carrier wavefunctions are quasi-localized over individual atoms or molecules, and carrier transport is by hopping from site to site which causes the mobility to be exceedingly poor. Any hopping transport can be viewed as a sequence of fast hops from one quasi-localized state to another, separated by exponentially long waiting periods at each localized state [102]. During the waiting period, the electron velocity is zero and the D'yakonov-Perel' mechanism is inoperative. This mechanism rotates the spin only during hopping and the amount of rotation is proportional to the displacement of the electron. Since the hopping distance is very short, it is unlikely that a complete spin precession will occur over such a short distance especially for organics which have very weak spin-orbit interaction. A rough estimate shows that in order to obtain a $\sim 180^o$ rotation during a hop (i.e. for a spin to flip during a hop), the spin-splitting energy due to spin-orbit coupling should be ~0.4 eV which is of the same order as many III-V semiconductors (e.g. GaAs, GaSb, InAs)[9]. Such a large spin splitting is unlikely to occur in organics which inherently has weak spin-orbit coupling. A similar argument regarding the ineffectuality of D'yakonov-Perel' mechanism in organics has been presented in [101]. We have also proved this experimentally by showing that spin relaxation rate in organics ($Alq_3$) is *not* proportional to mobility. We will come back to this point in section 3.2.

Thus the D'yakonov-Perel' mechanism is almost certainly not the dominant spin relaxation mechanism in organics. The Bir-Aronov-Pikus mode is usually important in bipolar transport (such as in spin-OLEDs) where both electrons and holes participate in conduction. But for unipolar transport where only one type of charge carrier, either electrons or holes, carries current, this mechanism is not present. That leaves the Elliott-Yafet and hyperfine interactions as the two likely mechanisms for spin relaxation. Which one is the more dominant may depend on the specific organic (e.g. its molecular structure, momentum scattering rate etc.). In the case of the $Alq_3$ molecule, from the observed dependence of spin relaxation rate on carrier mobility, temperature and transport-driving electric field, it is likely that the Elliott-Yafet mode is dominant at least at moderate-to-high electric fields [84, 100, 101]. This is probably because the carrier wavefunctions in this molecule are quasi-localized over carbon atoms whose naturally abundant isotope $^{12}C$ has no net nuclear spin and hence cannot cause significant hyperfine interaction. It has been experimentally demonstrated in [21] that hyperfine interaction does not play a major role in organic magnetoresistance in $Alq_3$.

However, in some other molecule where the carrier wavefunctions are spread over atoms with non-zero nuclear spin, it is entirely possible for the hyperfine interaction to outweigh the Elliott-Yafet mechanism. Recently it has been shown experimentally that hyperfine interaction plays a central role in the organic π conjugated polymer poly (dioctyloxy) phenylenevinyelene (DOO-

---

[9] Assuming a hopping distance of ~ 1 nm which is the average distance between two adjacent molecules. For D'yakonov-Perel' mechanism to be effective, this distance has to be of the order of the spin precession length, which is given by $h/2\sqrt{2m\Delta_{SO}}$ where $h$ is Planck's constant, $m$ is free electron mass and $\Delta_{SO}$ is the spin splitting energy.

[103]   M. Governale and U. Zulicke, "Spin accumulation in quantum wires with strong Rashba spin-orbit coupling," *Physical Review B*, vol. 66, pp. 073311, 2002.

PPV) [85].

## *2.4 Measurement of spin relaxation length and time: spin valve devices.*

A standard method to extract spin relaxation length ($L_s$) and spin relaxation time ($\tau_s$) in a paramagnetic material is to perform a spin-valve experiment. A spin-valve is a trilayered construct, in which the paramagnetic material of interest is sandwiched between two ferromagnetic electrodes of different coercivities (Figures 1b and 4a). Unlike giant magnetoresistive devices, these ferromagnets are not magnetically coupled with each other. As a result, their magnetizations can be independently controlled by a global magnetic field. One of these ferromagnets acts as spin injector *i.e.* under an applied electrical bias it injects spins (from the quasi-Fermi level) into the paramagnet. Electrical spin injection from a ferromagnet has been discussed before. The second ferromagnet also provides unequal spin-up and spin-down DOS at the Fermi level and preferentially transmits spins of one particular orientation. For the simplicity of discussion, let us assume that the detector ferromagnet transmits (blocks) completely the spins which are parallel to its own majority (minority) spins or the magnetization of the detector. The transmission probability (*T*) of an electron through the detector is then proportional to $cos^2 \theta/2$ where *θ* is the angle between the spin orientation of the electron arriving at the detector interface and the magnetization of the detector ferromagnet [3]. This means that if (*a*) the magnetizations (*M*) of the ferromagnets are parallel, (*b*) the injector injects majority spins and (*c*) there is no spin flip or spin relaxation in the spacer or the interfaces, the transmission coefficient should be unity (since $\theta = 0$) which will result in a small device resistance. Similarly, when the magnetizations are antiparallel (and other two conditions as before), the transmission coefficient should be zero (since $\theta = \pi$) and in this case one should observe large device resistance. In an actual device, there is always some spread in the value of *θ* since different electrons undergo different degrees of spin relaxation before arriving at the detector interface and this makes *θ* different for different electrons. As a result, ensemble averaging over all the electrons makes the resistance finite even when the magnetizations of the two ferromagnetic contacts are antiparallel. Nonetheless, if the length of the ferromagnetic spacer is smaller than the spin relaxation length, then the resistance is measurably larger when the contacts are in antiparallel orientation than when they are in parallel orientation. It follows that, if one of the ferromagnets preferentially transmits (either injects or detects) minority spins, then resistance of the spin valve will be large (small) in parallel (antiparallel) configuration. This is sometimes referred to as the "inverse spin valve" effect.

In a spin-valve experiment, resistance of the device is measured as a function of the applied magnetic field ($H$). The change in resistance between the parallel and antiparallel configurations allows one to extract the spin relaxation length and time in the paramagnetic spacer. For the following discussion, we will assume a regular spin valve effect i.e. low (high) resistance in the parallel (antiparallel) configuration. We will also assume that the coercive fields of the ferromagnets are given by $|H_1|$ and $|H_2|$ with $|H_1| < |H_2|$.

First, the device is subjected to a strong magnetic field (say $H_{sat} \gg |H_2|$) which magnetizes both the ferromagnets along the direction of the field and the resistance of this device is $R_P$ where the subscript ($P$) indicates "parallel" magnetization configuration. Next, the field is decreased, swept through zero and reversed. At this stage, when the magnetic field strength ($H$) just exceeds $|H_1|$ in the reverse direction (i.e. $-|H_2| < H < -|H_1|$), the ferromagnet with the lower coercivity (i.e. $|H_1|$) flips magnetization. Now the two ferromagnetic contacts have their magnetizations antiparallel to each other and the device resistance is $R_{AP}$ where the subscript ($AP$) indicates "antiparallel" magnetization configuration. For a regular spin valve effect as described above, at $H = -|H_1|$ a jump (increase) in the device resistance should occur implying $R_{AP} > R_P$. As the magnetic field is made stronger in the same (i.e. reverse) direction, the coercive field of the second ferromagnet will be reached ($H = -|H_2|$). At this point the second ferromagnet also flips its magnetization direction, which once again places the two ferromagnets in a configuration where their magnetizations are parallel. Thus, the resistance drops again to $R_P$ at $H = -|H_2|$. Therefore, during a single scan of magnetic field from $H_{sat}$ to $-H_{sat}$, a spin valve device shows a resistance peak between the coercive fields of the two ferromagnets (i.e. between $-|H_1|$ and $-|H_2|$). Similarly if the magnetic field is varied from $-H_{sat}$ to $H_{sat}$, an identical peak is observed between $|H_1|$ and $|H_2|$. The spin-valve response is pictorially explained in Figure 4b. The relative change in device resistance between the parallel and antiparallel configurations is the so called spin-valve peak and is expressed as $\Delta R/R = (R_{AP} - R_P)/R_P$.

For the special case when the spacer is a tunnel barrier, this ratio is expressed by the famous Julliere formula [104][10]:

$$\frac{\Delta R}{R} = \frac{R_{AP} - R_P}{R_P} = \frac{2P_1P_2}{1 - P_1P_2} \qquad (5)$$

where $P_1$ and $P_2$ are the *spin polarizations of the density-of-states (DOS) at the Fermi level* of the two ferromagnetic electrodes. Julliere's model assumes (a) spins are conserved (no spin reorientation/flip occurs) during tunneling and therefore tunnelings of ↑ and ↓ spin electrons can be viewed as two independent processes and (b) tunneling probability is spin independent and *conductance for a particular spin species* (↑/↓) *is solely determined by the appropriate density of states at the two ferromagnets*. The quantities $P_1$ and $P_2$ are often equated to the spin polarizations of the tunneling current which are independently determined via Meservey-Tedrow experiments using alumina tunnel barrier [88]. It is to be noted that the spin polarization of the tunnel current not only depends on the DOS but also depends on tunnel probability which is barrier-dependent and may be different for different electronic states in the ferromagnet. The importance of the barrier has been shown in [90]. According to this report, Co exhibits a negative spin polarization of tunneling electrons for $SrTiO_3$ barrier but a positive spin polarization for alumina barriers.

Julliere's formula can be extended for thicker paramagnetic spacers in which spin transport occurs via drift-diffusion or multiple hopping instead of direct tunneling between the contacts. However, the injector and detector interfaces are assumed to have a tunneling (Schottky) barrier, which occurs in many metal/organic interfaces. In this situation, spin polarized electrons are injected via the tunnel barrier into the paramagnet. Spin injection via thermionic emission is rather inefficient as described before. Let us assume that the spin polarization of the injected carriers is $P_1$ which can be determined exactly by the Meservey-Tedrow technique. Once in the paramagnet, these carriers will drift and diffuse (or hop) towards the detector contact, under the influence of a transport-driving electric field. The spin polarization of the injected carriers ($P$) decreases with distance $x$ (measured from the injection point) and this spin relaxation process can be modeled as $P(x) = P_1 exp(-x/L_s)$ where $L_s$ is the spin relaxation (or diffusion) length as described before. Exponential decay of spin polarization with distance has also been confirmed experimentally [98]. Thus when the carriers arrive at the detector interface, their spin polarization becomes $P(d) = P_1 exp(-d/L_s)$, where $d$ is the distance between the injector and the detector. Now we apply the Julliere formula on the tunnel barrier at the detector interface which separates the two spin polarizations $P(d)$ and $P_2$. In this case, using equation (5) we have

$$\frac{\Delta R}{R} = \frac{2P(d)P_2}{1 - P(d)P_2} = \frac{2P_1 P_2 \exp[-d/Ls]}{1 - P_1 P_2 \exp[-d/Ls]} \tag{6}$$

---

[10] Julliere's original paper (cited above), derives a relative change in conductance between the parallel and antiparallel configurations:
$(G_P - G_{AP})/G_P = 2P_1P_2/(1 + P_1P_2)$

This is the so-called "modified Julliere formula" which is widely used to estimate $L_s$. The physical model described above is depicted in Figure 4 (c).

In almost all applications of this model, the quantities $P_1$ and $P_2$ are taken directly from literature [88] and may not represent the exact values of spin polarization relevant for a particular experiment. For purely organic or organic-inorganic hybrid barriers, spin polarizations of the ferromagnets have been found to be less [91, 98] than the tabulated values [88]. Further, any surface contamination of the ferromagnets can also reduce $P_1$ and $P_2$. To include these effects, $P_1$ and $P_2$ can be replaced by $\alpha_1 P_1$ and $\alpha_2 P_2$ respectively where $\alpha_1$, $\alpha_2 < 1$. To determine $\alpha_1 P_1$ and $\alpha_2 P_2$ one needs to carry out spin dependent tunneling [89], muon spin rotation [105] or two photon photoemission experiments [106]. It is straightforward to show that $L_{s,actual} > L_{s,Julliere}$ where $L_{s,actual}$ ($L_{s,Julliere}$) is the $L_s$ value without (with) the assumption $\alpha_1 = \alpha_2 = 1$. Since $\alpha_1$ and $\alpha_2$ are generally unknown, the estimated $L_s$ (i.e. $L_{s,Julliere}$) provides a lower bound of the actual value and should be interpreted accordingly. In spite of this inherent limitation, this model provides valuable insight and can be used to obtain a rough estimate of $L_s$. Dependence of $L_s$ on temperature and bias can be used to shed light on the spin dynamics in the paramagnet.

## 3. Spin injection and transport in organics: spin valve experiments

Most studies in organic spintronics revolve around the so-called "organic spin valve" devices in which an organic material is contacted by two ferromagnetic electrodes (spin injector/detector). The organic layer is generally thick enough to avoid any coupling between the two ferromagnetic electrodes. The spin valve response ($\Delta R/R$) is measured as a function of temperature and bias. Julliere formula, modified for diffusive/hopping transport (equation 6), is employed to estimate the spin relaxation length (or more precisely, its lower limit). Then using the mobility of the carriers it is possible to estimate the spin relaxation time. Bias and temperature dependence of spin relaxation length and time indicate the dominant spin relaxation mechanism in organics. In this section we will discuss some of these studies.

It is to be noted that LSMO (lanthanum strontium manganate, $La_{1-x}Sr_xMnO_3$ with $0.2 < x < 0.5$) has been used as spin injector/detector in most of the studies of organic spin valve. LSMO is a half-metallic ferromagnet and acts as an excellent spin injector/detector due to near 100% spin polarization at low temperatures [107]. One advantage is that unlike ferromagnetic metals such as iron, nickel or cobalt, LSMO films are already stable against oxidation [108] and therefore immune against any possible degradation of surface magnetization or spin injection efficiency due to oxide formation at the interface. LSMO/organic interface is also likely to alleviate the conductivity mismatch problem and facilitate spin injection (see equation 3 and related discussion). On the downside, the spin polarization of LSMO decreases with temperature and poor at room temperature [109]. Thus, for room temperature organic spintronic devices, traditional ferromagnets with large and almost temperature independent spin polarizations (with a natural Schottky interface barrier) may be a better choice [96]. However, room temperature operation with LSMO electrodes has been reported recently [110].

### *3.1 Organic thin films.*

(*i*) *Sexithienyl ($T_6$) thin films*:

The first organic spin valve device was reported in 2002 [11, 12]. In this work, a thin film of an organic semiconductor sexithienyl ($T_6$) was used as the spin transport material. This material is a π-conjugated rigid rod oligomer (Figure 5) which, because of its relatively high values of field effect mobility (~$10^{-2}$ cm$^2$V$^{-1}$s$^{-1}$), on/off current ratio and switching speed, is an attractive channel material for organic field effect transistors [111-113]. Organic light emitting diodes based on $T_6$ have also produced polarized electroluminescence [114].

The spin valve device reported in [11] has a planar geometry in which two planar LSMO electrodes were separated by a $T_6$ spacer layer (Figure 5). It is to be noted that the $T_6$ film did not show any intrinsic organic magnetoresistance effect up to 1 Tesla [11]. The LSMO contacts exhibit intrinsic magnetoresistance of ~10-25%. However since the overall device resistance is primarily determined by the $T_6$ region whose resistance is approximately six orders of magnitude larger than that of LSMO, any observed magnetoresistance cannot accrue from LSMO. Further, the device resistance scales linearly with the thickness of the $T_6$ layer and therefore interface resistance is unlikely to contribute much in the overall device resistance.

The device resistance showed a strong dependence on magnetic field. It is to be noted that since both LSMO contacts were nominally identical, they did not allow independent switching of magnetizations as required in the case of a spin-valve. However, one could still change the magnetizations from random (zero field) to parallel (at high field). In the absence of any magnetic field, i.e. when the LSMO electrodes had random magnetization orientations, the device resistance was high. When a sufficiently large (saturation) magnetic field of 3.4 kOe was applied, the LSMO electrodes acquired magnetizations that were parallel to each other and the device resistance dropped. Maximum resistance change of $\sim 30\%$ was observed when the width of the $T_6$ spacer was $\sim 140$ nm. For larger spacer widths, the amount of resistance change decreased and disappeared beyond $\sim 300$ nm (equation 6). The observed magnetoresistance can be explained by invoking the spin valve effect described earlier and indicates spin polarized carrier injection and transport in $T_6$. The spin relaxation length ($Ls$) and spin relaxation time $\tau_s$ in $T_6$ was estimated to be $\sim 200$ nm and $\sim 1\mu$s respectively at room temperature. Interestingly, the resistance change was immune to the applied bias at least in the range of $0.2 - 0.3$ MV/cm as reported in [11]. However, in other organics (e.g. $Alq_3$; see later), it has been found that the spin-valve signal is strongly sensitive to applied bias and falls off rapidly with increasing bias.

Figure 5 shows an *inverse* spin valve effect observed in a *vertical* $T_6$ spin valve at 40K [14]. Unlike previous case, two different ferromagnets have been used (LSMO and Co), which allow independent switching. An alumina barrier has been employed to facilitate spin injection. Similar inverse spin valve effect has also been observed for other organics (e.g. $Alq_3$) as we discuss below.

*(ii) Tris 8-hydroxyquinoline aluminum ($Alq_3$) thin films*:

Arguably, one of the most studied organic materials in spintronics is $Alq_3$ which falls in the category of small molecular weight organic compounds. The molecular structure is shown in Figure 6 (*i*). This is an *n*-type organic which is frequently used in green organic light emitting diodes. The first study on this material [108] reported a vertical spin valve structure in which a thin film of $Alq_3$ (of thickness in the range 130-250 nm) is sandwiched between LSMO and cobalt contacts which serve as spin injector/detector (Figure 6 (*ii*)). Spins are injected from one of the ferromagnets and these spins undergo relaxation as they travel through the organic layer and are finally detected at the other ferromagnet. The effective thickness of the organic layer is smaller than the deposited thickness since the top cobalt layer, when deposited on $Alq_3$ tend to diffuse into the soft organic and forms a so-called "ill-defined" layer. The typical thickness of such ill-defined layer is ~ 100 nm [108, 115]. Such interdiffusion can be minimized by growing a thin tunnel barrier (e.g. of alumina) on $Alq_3$ prior to the deposition of the top ferromagnet or by cooling the substrate during deposition [116, 117]. Magnetic hysteresis loops reveal that the coercivities of LSMO and cobalt contacts are 30 Oe and 150 Oe respectively.

Magnetoresistance measurements of the spin valve device show resistance peaks between these coercivity values (Figure 6 (*iii*)). Interestingly, the sign of the spin valve peak is *negative*; i.e. the device resistance is *low* when the magnetizations are antiparallel and *high* when they are parallel. Same inverse spin valve effect was also observed for Fe/$Alq_3$/Co [115] and Ni/$Alq_3$/Co [118] devices.

From Figure 6 (*iii*), the spin valve signal (defined in equation 5) at 11 K is given by $(R_{AP} - R_P)/R_P \sim -0.3$. Using the modified Julliere formula (equation 6), it is possible to estimate the spin relaxation length $L_s$ in $Alq_3$ thin films. Assuming 100% spin polarization for LSMO [108], $-42\%$ spin polarization for cobalt [88, 108] and thickness of spin transport layer $d \sim (130 - 100)$nm $= 30$ nm, we obtain spin relaxation length $L_s \sim 45$ nm at 11 K. Similar values of $L_s$ have been reported by other studies such as [117]. However, as discussed before, Julliere's model does not account for possible loss of spin polarizations at the interfaces and the $L_s$ value derived from this model must be viewed as a lower bound rather than an exact value.

*Temperature dependence of the spin valve signal*:

The spin valve signal in Co/$Alq_3$/LSMO devices decreases with temperature and is almost non-existent at room temperature (Figure 6 (*iv*)). This according to equation (6), can happen due to two reasons: (1) enhanced spin relaxation rate in $Alq_3$ at elevated temperatures [108] (i.e. reduced $L_s$) and/or (2) reduced bulk spin polarization ($P_1$) of the LSMO electrode at elevated temperatures [109], since LSMO has a relatively low Curie temperature of $\sim 325\ K$ [110]. The bulk spin polarization of cobalt ($P_2$) is expected to be significant even at room temperature due to relatively high Curie temperature of $\sim$ 1150 $^o$C. Therefore if LSMO is replaced by a ferromagnetic metal (e.g. iron [115]) which has much larger Curie temperature ($\sim$ 768 $^o$C) than LSMO, one expects the spin valve signal to be present even at room temperature. However, spin valve signal still decreases with temperature and vanishes at a much smaller temperature ($\sim$ 90 K) [115]. This apparently indicates that spin relaxation rate in $Alq_3$ is temperature dependent. However, spin ½ photoluminescence detected magnetic resonance (PLDMR) experiments show that spin relaxation rate in $Alq_3$ is actually temperature *independent* [109, 110]. To resolve this issue it is necessary to understand that the surface spin-polarization of the ferromagnet (instead of bulk spin polarization) is mainly responsible for spin injection, and has much stronger temperature dependence than bulk. It is known that the surface magnetization of transition metal ferromagnets have a stronger temperature dependence than the bulk magnetization and this dependence is also related to the material grown on the ferromagnetic film [117, 119]. An improved fabrication of the ferromagnetic/organic interfaces is therefore necessary to obtain room temperature operation. Indeed, in carefully constructed Fe/$Alq_3$/Co devices with significantly reduced interdiffusion and surface roughness, the spin valve response has been found to persist up to room temperature (Figure 7a), which certainly bodes well for organic spintronics [116, 117]. For LSMO based spin valves, insertion of a tunnel barrier at the organic/ferromagnet interface has also resulted in room temperature operation [110] (Figure 7b). At present, the room temperature signal is very weak ($<$ 1%, Figure 7) and requires further improvement of the interface spin polarization. As discussed before, at higher temperatures, carrier injection may be dominated by thermionic emission, resulting in poor spin injection efficiency.

*Bias dependence of the spin valve signal:*

As shown in Figure 8a, for the LSMO/$Alq_3$/Co device, the spin valve signal decreases with bias. Again, according to equation (6) this can happen due to several reasons such as (*a*) increased spin relaxation rate (and therefore smaller $L_s$) in $Alq_3$ at high fields, (*b*) increased electron-magnon scattering (and hence smaller spin polarization) in the LSMO electrode at higher currents [16, 109, 116] and (*c*) change in the spin polarization of cobalt with bias [16, 109, 116]. The rate of decrease is asymmetric with respect to bias polarity even though the current-voltage characteristics (Figure 6b) are almost symmetric [108, 109]. This indicates that the spin injection/detection characteristics are different for Co/$Alq_3$ and LSMO/$Alq_3$ interfaces. Similar decrease in spin valve signal with bias has also been observed when both ferromagnets are transition metals (e.g. Ni and Co [118] or Fe and Co [116]). If enhanced spin scattering in $Alq_3$ at higher bias is responsible for this effect then it indicates that Elliott-Yafet mechanism is the dominant spin relaxation mechanism in this material [84].

*The sign problem:*

Unfortunately, understanding of the sign of spin valve response ($\Delta R/R$) is yet fragmentary. For example, Fe/$Alq_3$/Co devices show both negative spin valve ($R_{AP} < R_P$) [109, 115] and positive spin valve effect ($R_{AP} > R_P$) [117]. Similar behavior has also been observed in Ni/$Alq_3$/Co nanowire spin valves [118]. Inverse spin valve effect is regularly observed in LSMO/$Alq_3$/Co junctions [108, 110, 120, 121]. However, tunneling magnetoresistance measurements with $Alq_3$ barrier typically show positive spin valve effect [91].

According to equation 6, sign inversion of $\Delta R/R$ indicates sign inversion of $P_1$ or $P_2$. For Co/$Alq_3$ or Co/$Al_2O_3$/$Alq_3$ junctions, it has been shown that Co injects majority spins (i.e. spin moments parallel to magnetization) into $Alq_3$ i.e. $P_1 > 0$ [91]. LSMO has only majority states at the Fermi level and therefore tunneling spin polarization $P_2 > 0$, regardless of the nature of the insulating barrier [88]. Therefore a positive $\Delta R/R$ is expected but a negative $\Delta R/R$ is often observed.

The inverse effect is mostly observed in thick organics where transport occurs via drift-diffusion or multiple hopping instead of direct tunneling between the contacts. In such cases other effects may effectively invert the spin polarization of a ferromagnet [118, 122]. As mentioned before, during the fabrication of *vertical* spin valve devices with organic spacers, ferromagnetic atoms or clusters tend to penetrate into the $Alq_3$ layer upon deposition [108, 123, 124]. The penetrated ferromagnetic atoms and those at the interface react chemically with the organic to form a complex [123, 125]. Additionally, the ferromagnetic property of the top layer is also found to be influenced by the morphology of the underlying organic [117, 126]. While the implications of these issues on spin injection is not fully understood, it has been suggested that the change in the sign of the spin valve may occur due to resonant tunneling of the carriers through the impurity states that originate during the deposition of the top ferromagnetic metal on "soft" $Alq_3$ [118, 122]. Another possibility is the existence of pinholes in the organic layer which can also invert the spin valve signal [127]. We note that with devices in which special caution was taken to avoid ferromagnet interdiffusion in organics (e.g. substrate cooling or barrier deposition [91, 116, 117]), sign of the spin valve is generally positive. Devices in which significant amount of ferromagnetic species interdiffuses into the organic to form a so-called "ill-defined" layer, typically show a negative spin valve effect [108, 115]. An exception is [110] which reports efficient spin injection at room temperature, which most likely accrues from the presence of the tunnel barrier. Still, this device shows an inverse spin valve effect. Direct characterization of the spin polarization of the injected current at the interfaces may offer more insight in this puzzling feature.

The presence of a tunnel barrier on top of the "soft" organic layer is beneficial due to several reasons. First, as mentioned above, this layer prevents interdiffusion of the top ferromagnetic contact during deposition and results in a well-defined interface. Second, it has been reported that in the absence of the barrier, molecular $Alq_3$ anions are formed at the interface due to charge transfer from the transition metal ferromagnet [128]. The tunnel barrier prevents this charge transfer and helps preserve the $Alq_3$ electronic structure within the interface region [128]. Third, it is known from the wisdom of conventional organic devices that such a buffer layer generally improves device performance [110, 129-131] by reducing the formation of interfacial trap states and fourth, the barrier can also enhance spin-injection efficiency since it ameliorates the notorious conductivity mismatch problem [96]. Finally, the quality of the ferromagnet film deposited on alumina layer is significantly better than the one on bare $Alq_3$ [131]. This high quality ferromagnetic contact is essential for spin injection. For a direct Co/$Alq_3$ junction, the interfacial Co layer (~ 3.5 nm) is not ferromagnetic at room temperature and it has been suggested that this layer may act as a spin scattering agent and adversely affect spin injection [131]. Similar "magnetically dead" layers exist for Fe/$Alq_3$ junctions [117] as well as LSMO-RRP3HT junctions [132]. However in these studies the authors suggested that this layer can act as a spin selective barrier and help avoid conductivity mismatch problem resulting in efficient spin injection [117, 132]. Further studies will be necessary to identify the exact role played by this layer.

(*iii*) *Other organics.*

While $Alq_3$ is probably the most studied spin-transport material in organic spintronics [14], other small molecular weight and polymeric organics have also been investigated by several groups [16, 85, 98, 109, 120, 121, 132-137]. Room temperature positive spin valve effect was found in $Fe_{50}Co_{50}$/RR-P3HT [11]/$Ni_{81}Fe_{19}$ spin valve [134]. Spin relaxation length of ~ 62 nm has been reported for this system [134]. Similar magnetoresistance study with LSMO electrode has been reported in [120, 132, 133]. The authors estimated a spin relaxation length of ~ 400 nm and spin relaxation time of 7 ms at low temperatures and bias. Magnetoresistance up to 20% has been reported in π-conjugated molecular pyrrole derivative 3-hexadecyl pyrrole (3HDP) at room temperature [135]. Spin relaxation length in rubrene has been found to be ~ 13.3 nm at low temperature [98]. Unlike most of the other studies which use a spin valve configuration, this measurement has been performed by a spin-polarized tunneling technique. The presence of an $Al_2O_3$ seed layer enhances the spin transport properties due to improved growth of organic semiconductor. Ref. [127, 139] reported rubrene spin valves and confirm the importance of an interfacial barrier for efficient spin injection. Spin valve device with TPP (tetraphenyl porphyrin) spacer and LSMO/Co contacts has been demonstrated in [121]. These devices also show a negative spin valve peak which gradually vanishes with increasing temperature and bias, consistent with the observations made on $Alq_3$ [108]. Many other organics have also been studied recently, which include $C_{60}$ [16], PPV[85], CVB [109], NPD [109], CuPc [116], PTCDA [116], $CF_3$-NTCDI [116], pentacene [140] and BTQBT [140]. Unlike most other approaches, ref. [140] fabricated a planar spin valve (similar to the original work [11]) to avoid the "ill-defined layer" at the interface.

---

[11] RR-P3HT: regioregular poly(3-hexylthiophene) also exhibits the so-called organic magnetoresistance (OMAR) effect in which a significant magnetoresistance is observed even when the electrodes are non-magnetic.

[138] O. Mermer, G. Veeraraghavan, T. L. Francis, Y. Sheng, D. T. Nguyen, M. Wohlgenannt, A. Kohler, M. K. Al-Suti, and M. S. Khan, "Large magnetoresistance in nonmagnetic pi-conjugated semiconductor thin film devices," *Physical Review B*, vol. 72, pp. 205202, 2005.

*3.2 Organic nanowires.*

The experiments discussed above clearly demonstrate that spins can be electrically injected into an organic semiconductor and reasonably long spin relaxation length and time can be achieved. Even room temperature operation is possible with careful fabrication of the interfaces. However, they do not shed much light on the nature of the dominant spin relaxation mechanism in organics. Presence of LSMO electrode with a temperature and bias dependent spin polarization makes this analysis even more complicated. Lack of this knowledge motivated us to investigate spin transport in organics with a view to establishing which spin relaxation mechanism is dominant. Consequently, we focused on organic *nanowires* [100, 118] instead of standard two-dimensional geometries since comparison between the results obtained in nanowires and thin films can offer some insight into what type of spin relaxation mechanism holds sway in organic semiconductors.

Carriers in nanowires will typically have lower mobility than in thin films because nanowires have a much larger surface to volume ratio and hence carriers experience more frequent scattering from charged surface states in nanowires than they do in thin films. This scattering is *not* surface roughness scattering but rather Coulomb scattering from the charged surface states. The surface roughness scattering does not increase significantly in nanowires since the mean free path in organics is very small (fractions of a nanometer) and as long as the nanowire diameter (~50 nm) is much larger than the mean free path, we do not expect significantly increased surface roughness scattering. However, the Coulomb scattering from surface states increases dramatically. The Coulomb scattering is long range and affects carriers that are many mean free paths from the surface. As a result, nanowires invariably exhibit significantly lower mobility than thin films [15].

This mobility difference offers a handle to probe the dominant spin relaxation mechanism in organics. As explained earlier in section 2.2, the Elliot-Yafet spin relaxation rate is directly proportional to the momentum relaxation rate and hence inversely proportional to the mobility while the D'yakonov-Perel' rate is directly proportional to mobility. Hence, if we observe an increased spin relaxation rate in nanowires compared to thin films, then we will infer that Elliot-Yafet is dominant over D'yakonov-Perel'; otherwise, we will conclude that the opposite is true. We perform this test on $Alq_3$, since this material has weak hyperfine interaction [21] and Bir-Aronov-Pikus mode is not efficient since $Alq_3$ is primarily an electron transport material.

We carried out experiments that clearly showed the spin-valve effect in organic *nanowires* [100, 118]. These nanowire spin valve structures were synthesized using an electrochemical self-assembly technique which is a very commonly used approach for fabricating quasi-periodic arrays of nanowires or nanodots of a variety of materials (single or multilayered) on arbitrary substrates [141]. In this method an anodic alumina film containing well-regimented nanopores (Figure 9a) is synthesized by anodizing an aluminum foil under appropriate electrical bias. Next, desired material (metals, inorganic/organic semiconductors, molecules, nanotubes) are deposited or grown in these nanopores by employing various methods such as electrodeposition, pressure injection, evaporation, chemical vapor deposition and electrospraying. In general these pores are blocked at the bottom (aluminum/alumina interface) by a continuous alumina barrier layer. For transport experiments, this insulating layer needs to be removed prior to material deposition, via an isotropic etching technique. For fabricating the nanowire spin valve devices, we electrodeposit Ni at the bottom of the pore by applying a dc bias. Then $Alq_3$ was thermally evaporated on top of Ni followed by Co evaporation without breaking the vacuum. Finally gold wires were attached to the Co layer and Al substrate using silver paste. Figure 9b shows such an array of nominally identical nanowire spin valve structures. TEM image of a nanowire spin valve is shown in Figure 9c. The details of the fabrication process have been described elsewhere [15] and will not be repeated here. The coercivities of Ni and Co nanowires are significantly different and have been characterized extensively before [100, 142, 143]. This allows us to establish parallel and antiparallel magnetization configurations of the electrodes as required for spin valve operation. The ferromagnetic nanowires are magnetized along the longitudinal axis and parallel to the direction of current. Unlike other configurations reported in literature [80, 144], in this geometry there is no significant fringing field, transverse to the direction of current, that can potentially generate a local Hall voltage and mimic the spin valve signal.

The I-V characteristics is piecewise linear and almost independent of temperature [100] which supports the previously discussed transport model, in which carriers are first injected via tunneling through the interfacial barrier and then propagated by drift-diffusion or hopping from site to site [84]. No significant interdiffusion of the top Co layer into the underlying $Alq_3$ was observed, presumably due to the fact that the alumina template acts as a shadow mask [15].

It is important to note that the resistivities of the ferromagnetic nanowire electrodes are almost nine orders of magnitude smaller than the intermediate $Alq_3$ region and therefore we always probe the resistance of the $Alq_3$ region only and not the resistance of the ferromagnetic electrodes which are in series with the $Alq_3$ layer [15, 118]. Thus all features in the magnetoresistance and I-V plots accrue from the organic layer and has nothing to do with the ferromagnetic contacts. Consequently, if there are features originating from the anisotropic magnetoresistance effects in the ferromagnets, we will never see them in our measurements. Detailed description of the control experiments for nanowire geometry can be found in [15, 118].

These devices show both normal and inverse spin valve effects. The inversion of the spin valve signal is due to carriers resonantly tunneling through a localized defect or impurity state in the organic [118]. The normal spin valve response of these devices is shown in Figure 10 (*i*). The spin valve peaks are superimposed on a background magnetoresistance. Using Julliere formula, we were able to estimate the spin relaxation length ($Ls$) and the spin relaxation time ($\tau_s$) in nanowires, contrast them with the corresponding quantities measured in thin films, and thus determine the dominant spin relaxation mechanism in organics. We found that the spin relaxation rate in nanowires is about an order of magnitude larger than in thin films [100], which immediately suggests that the dominant spin relaxation mode is the Elliot-Yafet channel. This is consistent with our previous arguments regarding the weakness of D'yakonov-Perel' mode in organics. We also showed that the spin relaxation time ($\tau_s$) in $Alq_3$ can approach 1 second at a temperature of 100 K. This is the longest spin relaxation time reported in any nanostructure above the liquid nitrogen temperature (77 K). (Figure 10 (*ii*)). Here we would like to stress the fact that this estimate of $\tau_s$ is merely a worst case estimate. These values have been calculated by assuming 33% and 42% spin polarizations of Ni and Co respectively [88]. However, these numbers are valid for a clean interface, which is probably not applicable in this experiment, especially because the Ni electrode has been fabricated by electrochemical deposition and is prone to surface contamination. The actual spin polarization may be much less than the used values [145]. Further, we assume no loss of spin polarizations at the interfaces. These effects may be incorporated by the parameters $\alpha_1$ and $\alpha_2$ (equation 6 and following discussions) which will result in larger value of $L_s$ and hence a larger value of $\tau_s$.

We have also observed a surprising correlation between the sign of the spin-valve peak and the background magnetoresistance in $Alq_3$ nanowires [118]. We offered a possible explanation for this intriguing correlation and in the process showed that a magnetic field can increase the spin relaxation rate in organics, which is consistent with the Elliott-Yafet mechanism. It is consistent in two ways: (1) First, a magnetic field bends the electron trajectories bringing them closer to the surface of the nanowire, which decreases mobility and increases the Elliott-Yafet spin relaxation rate. (2) Second, a magnetic field causes "spin mixing" which increases the admixing of spin-up and spin-down states which exacerbates spin relaxation. The increase of spin relaxation rate in a magnetic field lends further support to our conclusion that the Elliott-Yafet mode is the dominant spin relaxation mechanism in $Alq_3$.

**4. Spin injection and transport in organics: Meservey-Tedrow spin polarized tunneling,**

**two photon photoemission and μSR experiments**

Thus numerous organic spin valve devices have been reported in literature which demonstrates the possibility of electrical spin injection, transport and detection in organic semiconductors. However, as discussed before, analysis of spin valve data underestimates $L_s$ since the actual values of surface spin polarizations $P_1$ and $P_2$ in Julliere formula are unknown and typical bulk values are used. Spin injection can also be demonstrated by alternate, more complex approaches such as spin polarized tunneling, two electron photoemission and muon spin rotation. In these methods we still need a ferromagnetic spin injector, but not a ferromagnetic spin detector. Spin detection is performed by alternative techniques. In this section we will briefly discuss these non-conventional methods.

*4.1. Meservey-Tedrow spin polarized tunneling.*

Ref. [98] employed Meservey-Tedrow spin polarized tunneling technique to determine the spin relaxation length $L_s$ of rubrene, which is a $\pi$ conjugated molecular semiconductor with chemical formula $C_{42}H_{28}$. In this experiment the device has a spin valve like structure but with only one ferromagnetic electrode. The other electrode is aluminum which is superconducting below 2.9 K. In presence of an in-plane magnetic field the superconducting aluminum acts like a spin detector since the quasi-particle DOS is Zeeman split. From the conductance vs. bias plots it is possible to calculate the spin polarization of the tunneling electrons that reach the aluminum interface. A review of this technique is available in [89].

Ref. [98] determined the tunneling spin polarization for various thicknesses of the rubrene layer with and without an additional tunnel (alumina) barrier. As the thickness increases, spin polarization of the current detected at the Al interface, decreases. From this analysis, $L_s$ in rubrene has been estimated to be ~ 13.3 nm. The presence of the alumina tunnel barrier significantly improves spin injection, as expected. This method of determining $L_s$, though direct and unambiguous, is not applicable at higher temperatures.

*4.2 Two photon photoemission spectroscopy*

*4.2.1. Background.*

Photoemission spectroscopy is among the most popular and versatile methods for studying solid surfaces and adsorbates [146]. In a typical photoemission process electrons are photoexcited from below the Fermi level to the vacuum level. The extracted photoemission spectrum contains energy distribution of the photoelectrons which provides quantitative information about the initial density of states. In particular "two photon photoemission" (TPPE) has attracted special attention as it allows simultaneous detection of occupied and unoccupied excited electronic states in a single measurement. It is a second order process where the photon energy has to be less than the work function of the material under study. The two photons termed as pump and probe photons can have same or different energies. The pump photon excites an electron from an occupied level lying below the Fermi level to an unoccupied intermediate level in between Fermi level and vacuum. From this intermediate level electron is photoemitted after absorbing the second photon (probe photon).

*4.2.2. Spin injection and transport studies by two photon photoemission (TPPE) spectroscopy.*

Ref. [106] has reported a spin resolved TPPE experiment that determines spin injection efficiency and spin diffusion length in a Co/ CuPc (copper phthalocyanine) heterojunction. In this study two 3.1 eV photons were provided by femtosecond pulsed laser beams with an adjustable time delay. The photon energy is chosen in such a way that a photoelectron can be emitted only if two photons are absorbed. Light penetration depth with this energy is significantly higher in Co (~22 nm) and CuPc (~530 nm) compared to the probing depth of photoemission, given by the inelastic mean free path of the optically excited electrons in CuPc (~1 nm). Spin polarized electrons from Co are first excited by the pump laser into an intermediate state which lies above the Fermi level (Figure 11). Part of these excited electrons will cross the ferromagnet-organic semiconductor interface. The spin polarized electrons lying above the lowest unoccupied molecular orbital (LUMO) of CuPc are excited by the probe laser which gives rise to photoemission spectra. Information about the spin injection efficiency is obtained from very thin coverage of the organic on the Co surface. The estimated spin injection efficiency from Co to CuPc was as high as 85% at room temperature. As expected, spin polarization of the photoemitted electrons decreased with increasing CuPc thickness. This dependence shows that spin relaxation length $L_s$ in CuPc is ~ 12.6 ± 3.4 nm at room temperature. It is expected that $L_s$ should increase at lower temperatures.

Ref. [116] reported spin valve effect in Fe/ CuPc (~100 nm)/Co devices. At 80 K, $\Delta R/R \sim 3\%$. Using bulk spin polarization values of 45% (Fe) and 42% (Co) [88], spin relaxation length $L_s$ turns out to be ~ 39 nm at 80 K. This reasonably good agreement with the TPPE experiment may be partially attributed to the high spin injection efficiency ($\alpha_1, \alpha_2 \approx 1$) at the ferromagnet/CuPc interfaces.

### 4.3. Low-energy muon spin rotation.

#### 4.3.1. Background.

Muon spin rotation (μSR)[12] technique is extensively used in condensed matter physics as an ultrasensitive probe of the internal magnetic fields of various materials. This technique is able to sense very weak magnetic fields of nuclear and atomic origin and can measure dynamic magnetic field fluctuations in the range of $10^4$ to $10^{12}$ Hz, depending on the size of the magnetic field at the muon site. Advances in this area have resulted in "slow muons" or "low energy muons" that are suitable for probing nanostructures [147]. For details of the μSR spectroscopy, the interested reader may refer to [147, 148] and the references therein.

Negative muons ($\mu^-$) are spin ½ particles with a lifetime of ~ 2.179 μs and they spontaneously decay into an electron and a neutrino-antineutrino pair. Due to their spins muons possess a magnetic moment $\mu_\mu$. So, when implanted in a well defined location within a solid, this magnetic moment couples to any magnetic field in its vicinity and serves as an extremely sensitive microscopic probe of the internal magnetic field of the solid. Detection of the evolution of the muon spin will therefore provide a detailed picture of the internal magnetism. Positive muons are generally used for the μSR experiments since negative muons have the deleterious effect of radiation damage and potential disintegration of the sample.

---

[12] Strictly speaking, "μS" stands for *Muon Spin* whereas "R" represents any of rotation/relaxation/resonance.

In a μSR experiment, spin-polarized (positive) surface muons are irradiated on a sample and they get implanted at various locations inside the sample. These implanted muon decays according to the equation

$$\mu^+ \rightarrow e^+ + \nu_\mu + \nu_e$$

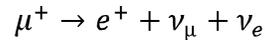

generating positrons, neutrinos and antineutrinos. This decay is anisotropic and the positron is emitted preferentially along the direction of the muon spin at the moment of decay. Detection of the anisotropic distribution of the positrons allows monitoring the temporal evolution of average direction of muon spin as a function of magnetic field and other variables. The time evolution of muon spin polarization is coupled to the magnetic environment of the muon and thus this technique allows access to the dynamics of the internal magnetic field of the sample.

*4.3.2. Measurement of spin diffusion length in organics by μSR spectroscopy.*

Ref. [105] reported the first direct measurement of spin diffusion length in a functional spin valve device using low energy μSR technique. The device schematic is shown in Figure 12 a. This device has the configuration of an organic *pn* diode where the *p* layer is made of TPD[13], a hole transport organic and the *n* layer is made of $Alq_3$ (200 nm), which is an electron transport material. Spin injection and detection is performed by FeCo and NiFe (17 nm) contacts. As seen before, an interfacial Schottky barrier facilitates spin-polarized injection and hence a lithium fluoride tunnel barrier has been fabricated at the interface of NiFe and $Alq_3$. The magnetic field (which aligns the injected spins) is applied parallel to the layers and perpendicular to the muon's initial spin direction and momentum. When spin polarized electrons are injected, the μSR spectrum is expected to change due to the interaction of the muon spins and the spins of the injected electrons. This change is used to determine the electron spin diffusion length $L_s$ in the organic $Alq_3$.

Figure 12b shows the typical depth profile which indicates that most of the muons are implanted well inside the organic layer $Alq_3$. At a temperature of 90 K, this corresponds to implantation energy of 6.23 keV. The ferromagnetic contacts were magnetized and saturated by application of an external magnetic field of ~ 100 mT which is much larger than the saturation magnetization (< 50mT). Then the magnetic field is reduced to a small value of ~ 29 mT. The μSR spectrum is first obtained with current on (electron spin injection) and then with current off (no spin injection). To confirm that the change in spectra is due to the injected spins and not due to the small external magnetic field, the following experiment was done. The magnetization of the injecting ferromagnet was reversed while the direction of the external magnetic field remained the same as before. In this case, the injected spins are mostly antiparallel to the external magnetic field and the μSR spectrum shifts in the opposite direction. This is a direct indication of current induced spin injection in organics and cannot be explained by intrinsic organic magnetoresistance which manifests even for nonmagnetic contacts. Analysis of this data reveals that the spin relaxation length $L_s$ in $Alq_3$ thin films is ~ 30 nm at ~ 10 K, which is in agreement with [108]. However, spin relaxation rate has been found to decrease with decreasing temperature, which contradicts the observation made in [109].

---

[13] TPD: N, N'-diphenyl-N-N'-bis(3-methylphenyl)-1,1'-biphenyl-4, 4'-diamine

## 5. Outlook and conclusion.

In this article we have briefly discussed the major spin transport experiments that have been performed in the last eight years on organic semiconductors. Initial studies reported low temperature operations but recently room temperature operation has been demonstrated [110, 117]. This undoubtedly makes organics a strong competitor for developing practical spintronic devices. The room temperature signal is still too weak to be useful in practical applications. However, there is enormous room for improvement. As mentioned before, one of the main attractive features of organics is that it allows chemical tuning of its physical properties such as HOMO-LUMO gap, injection barrier, mobility, spin-orbit coupling and hyperfine interaction [21, 85]. This is an area which is still largely untapped by the organic spintronics community. In future, organics may replace inorganic materials in certain niche applications such as flexible and inexpensive MRAMs and magnetic field sensors. Extremely long spin relaxation times have been reported in organics [78, 100] which bodes well for several other applications such as spin

39based classical and quantum computing and spin based organic light emitting diodes. Some of these applications are briefly discussed below.

(*a*) *Non-volatile memory and magnetic field sensors:* The most direct application of organic spintronics is likely to emerge from the area of flexible non-volatile memories. There is significant current interest in this area since these devices can be augmented with large area bendable sensor and actuator arrays to store the spatial distribution of the output data for a long time. Organic non-volatile memory transistors have already been demonstrated which have similar floating gate configuration as that of silicon flash memory transistors [149]. A natural extension of this idea is to realize organic based magnetic random access memories (OMRAM). Traditional inorganic MTJs (e.g. ferromagnet/alumina tunnel barrier/ferromagnet) can be grown on flexible plastic substrates [150]. However, the bending stress can have adverse effect on the tunneling barrier. A simpler alternative is to use (non-hybrid) organic semiconductors as tunnel barriers in MTJ devices. Recent studies of spin transport via organic tunnel barriers have shown change in magnetoresistance which is promising for this purpose [91]. Spin valve signal as high as 330% has been reported, albeit at low temperature [85]. OMRAM is non-volatile (like MRAMs and flash) and is expected to offer superior durability and faster access speeds than organic flash memories. However, large area assembly of such OMRAM cells has not been reported yet, even though it is achievable via self-assembly techniques [100]. Recently ref. [151] has reported a novel coexistence of electrical and magnetic bistabilty in $Alq_3$ based spin valves which enables storage of four distinct resistance states per cell. Ferromagnetic nanoparticles embedded in organic matrix shows large magnetoresistance [152-154], which can be harnessed in magnetic field sensor applications.

(*b*) *Spin based classical and quantum computing*: For both classical and quantum spin based computing, it is important to preserve the fidelity of the data during computation. Since data is encoded in the spin states, any unwanted coupling of spins with the environment during the computational cycle can potentially corrupt the stored data. The probability of such error during one computational cycle is given by

$$p = 1 - e^{\frac{-T}{T_s}} \qquad (7)$$

where $T$ is the duration of a computational cycle (typically the period of the clock that drives the computation) and $T_S$ is the spin relaxation time [155]. Clearly if $T_S$ can be made much larger than $T$, then $p$ will be approximately zero, implying low error probability. For the classical case, $T_S$ is equal to the spin flip time or the longitudinal spin relaxation time $T_1$. For the quantum case $T_S$ is generally equal to the transverse spin relaxation time $T_2$ (or $T_1$, whichever is shorter).

Now, we have seen that for $Alq_3$ nanowires $T_1$ time exceeds 1 second at a temperature of 100K [100]. Inorganic semiconductors have a similar $T_1$ time but only at a very low temperature of ~ few millikelvins [156]. This makes organics a preferred platform for classical spin-based computing. With $T_1 = 1$ second and for a 10 GHz clock ($T = 0.1$ nanosecond), according to the above equation, the classical bit error probability will be approximately $10^{-10}$ which may be sufficiently low to allow large scale fault tolerant spin-based classical computing. Further, for $Alq_3$ molecules the $T_2$ time is approximately 30 nanoseconds at room temperature and according to the above equation, the error probability is ~ 3% [157]. This is still low enough to allow fault tolerant quantum computing with the aid of error correction codes [158]. Additionally $Alq_3$ has special spin dependent optical properties (see below) that can be harnessed for easy qubit readout. Therefore organic semiconductors can supersede their inorganic counterparts in spin based computing applications – both classical and quantum mechanical.

(*c*) *Spin based organic light emitting diodes (spin-OLEDs)*: Long spin lifetime in organics can have a significant impact in "organic optospintronics" where spin effects can be fruitfully employed to boost the internal quantum efficiency of the organic light emitting diodes (OLED). It is often claimed that the worldwide OLED market will reach $ 15.5 billion (US) by 2014 [159], since they are inexpensive compared to the inorganic (semiconductor) LEDs, and can be produced on flexible substrates. An organic light emitting diode is basically a *p-n* junction where the *p* and *n* layers are made of organic semiconductors. Under forward bias, the electrons injected from the *n*-layer and the holes injected from the *p*-layer pair up to form excitons near the junction (assume, for simplicity of discussion, that the probability for this process is unity). An exciton can be either a singlet (spin $S = 0$) or a triplet (spin $S = 1$), depending on the spins of the electrons and holes [160]. The singlet exciton decays radiatively and rapidly, emitting a

photon but the triplet exciton decays non-radiatively and relatively slowly, emitting phonons (heat) rather than photons (light). Thus, the relative population of singlet and triplet excitons determines the quantum efficiency of the OLED. For unpolarized carrier injection, the probability of forming a triplet is 75%. Therefore at least 75% of electron-hole recombination events are wasted in heat, and the maximum quantum efficiency is limited to meager 25% [160].

This situation can be improved if spin polarized electrons and holes are injected in the OLED. It would then be possible to preferentially form singlets or triplets by controlling the spin polarizations of the injected carriers. Using this method, the relative population of the singlet and hence the internal quantum efficiency can reach a theoretical maximum of 50% [160], leading to brighter, energy-efficient and inexpensive OLEDs. For this to happen it is necessary to ensure that the exciton formation and radiative recombination rates are much larger than the spin relaxation rate. Thus for this purpose, long spin relaxation times are desirable and are indeed achievable in optically active organics like $Alq_3$.

It has been reported that spin-polarized injection increases the singlet-to-triplet ratio in organic semiconductor polymer MEH PPV and increases the electroluminescence intensity [161]. In this work spin polarized holes and unpolarized electrons were injected in an OLED structure which resulted in ~ 18% increase in the electroluminescence efficiency compared to the case when both type of carriers are unpolarized. Room temperature spin injection in electron and hole carrying organics has also been demonstrated recently [116] which is encouraging for spin-OLED research.

Thus it is possible to realize several spintronic applications using organics. All these applications exploit the inherent material properties of organics, such as flexibility, low cost processing and lightweight and therefore seem to be commercially viable. Further, to completely exploit the structural flexibility of the organic materials, it would be necessary to realize all-organic spin valves which rely on organic molecular magnets [162, 163] as spin injectors and detectors. Unlike inorganic magnets, properties of these materials are chemically (and even optically) tunable and they can be fabricated at low temperature [164]. This can lead to a well-defined interface with the paramagnetic organics and improve spin injection. These materials have semiconducting to insulating conductivity which can help circumvent the conductivity mismatch problem often encountered in spin injection experiments. These possibilities are largely unexplored at this point but if realized, can lead to completely flexible non-volatile memory chips and efficient displays.

## 6. Acknowledgement.

The authors gratefully acknowledge financial support from the Natural Sciences and Engineering Research Council (NSERC), TRLabs and the University of Alberta.

**Figures.**

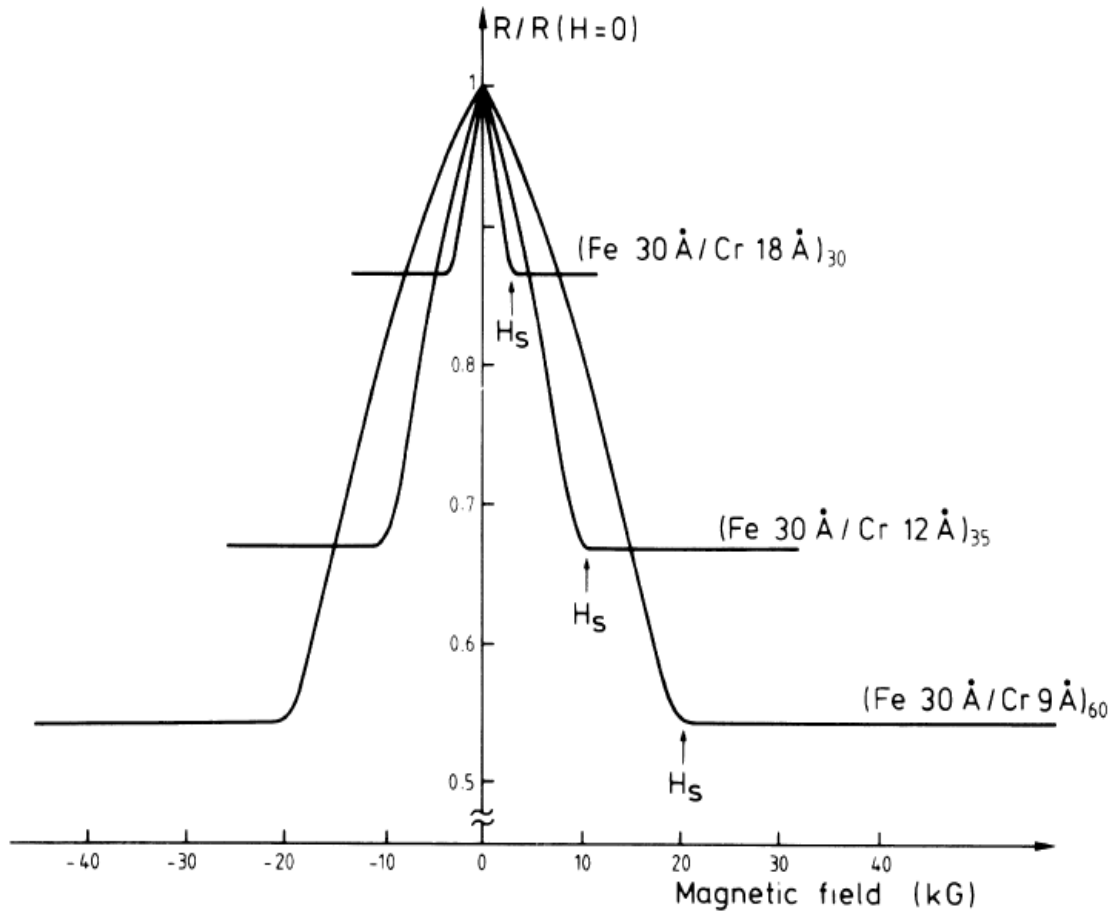

**Figure 1.** *Spintronics in data storage and information processing*. **(a)** Giant magnetoresistance (GMR) observed in Fe/Cr multilayers at 4.2 K (after Baibich et al., [22]). Change in device resistance with magnetic field is several orders of magnitude larger compared to anisotropic magnetoresistance (AMR) effect observed in ferromagnets. The traditional AMR read heads in hard drives have therefore been replaced by GMR heads, leading to exponential increase in areal data storage density.

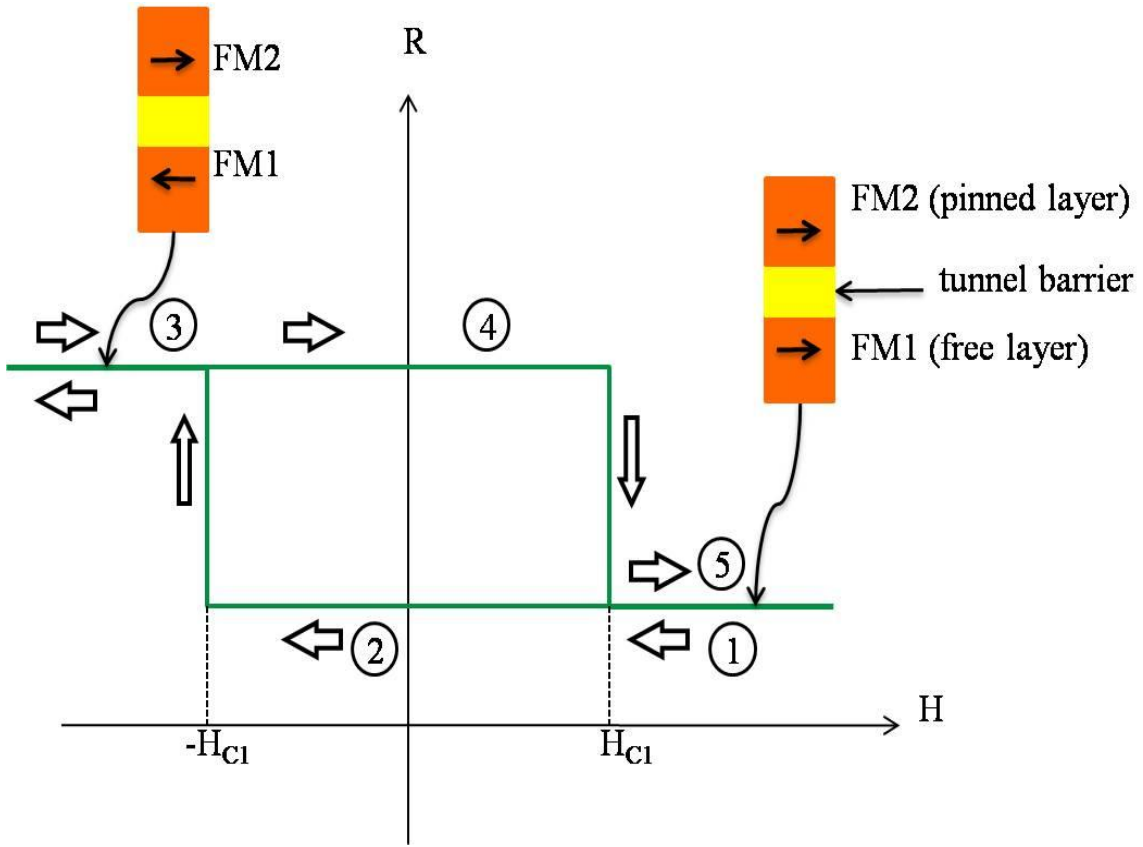

**Figure 1 (contd.)** *Spintronics in data storage and information processing.* **(b)** Hysteretic resistance ($R$) vs. magnetic field ($H$) characteristic of MTJ memory cell. The magnetization of the top ferromagnet (FM2) is fixed (or "pinned") whereas the magnetization of the bottom ferromagnet (FM1) can be changed. The coercivity of FM1 is $H_{c1}$. By varying the external magnetic field ($H$) the relative magnetization orientations can be changed and *bistable resistance state* can be realized. The arrows and numbers indicate the direction of scan of $H$. These bistable states can be used to store logic bits in magnetic random access memory (MRAM).

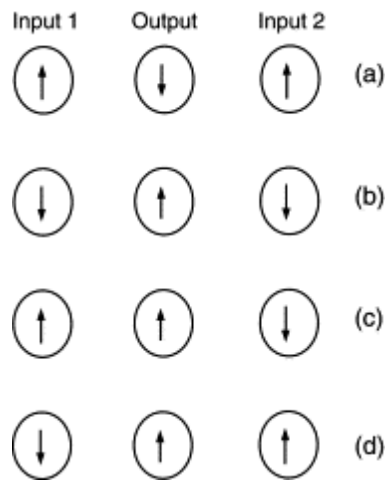

**Figure 1 (contd.)** *Spintronics in data storage and information processing.* **(c)** All-spin realization of a single NAND gate (after Bandyopadhyay et al. [10]). The edge spins indicate the input bits and the middle spin is the output bit. Logic 1 is encoded by upspin (↑).

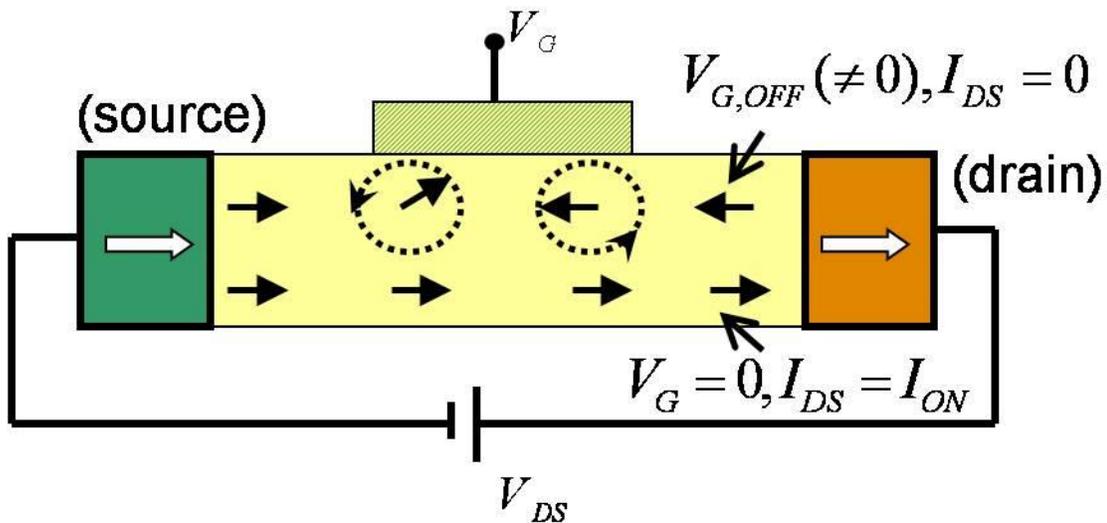

**Figure 1 (contd.).** *Spintronics in data storage and information processing.* **(d)** Schematic of a spin field effect transistor (adapted from [17, 47]). When the source and drain magnetizations are parallel and gate voltage $V_G = 0$, the channel resistance is low (similar to the spin valve effect in Figure 1b) and the transistor is on. Applying a particular gate voltage $V_{G,off}$, rotates the spins by an angle $\pi$ via Rashba spin-orbit interaction. These spins are now blocked by the drain resulting in high device resistance and the transistor is off.

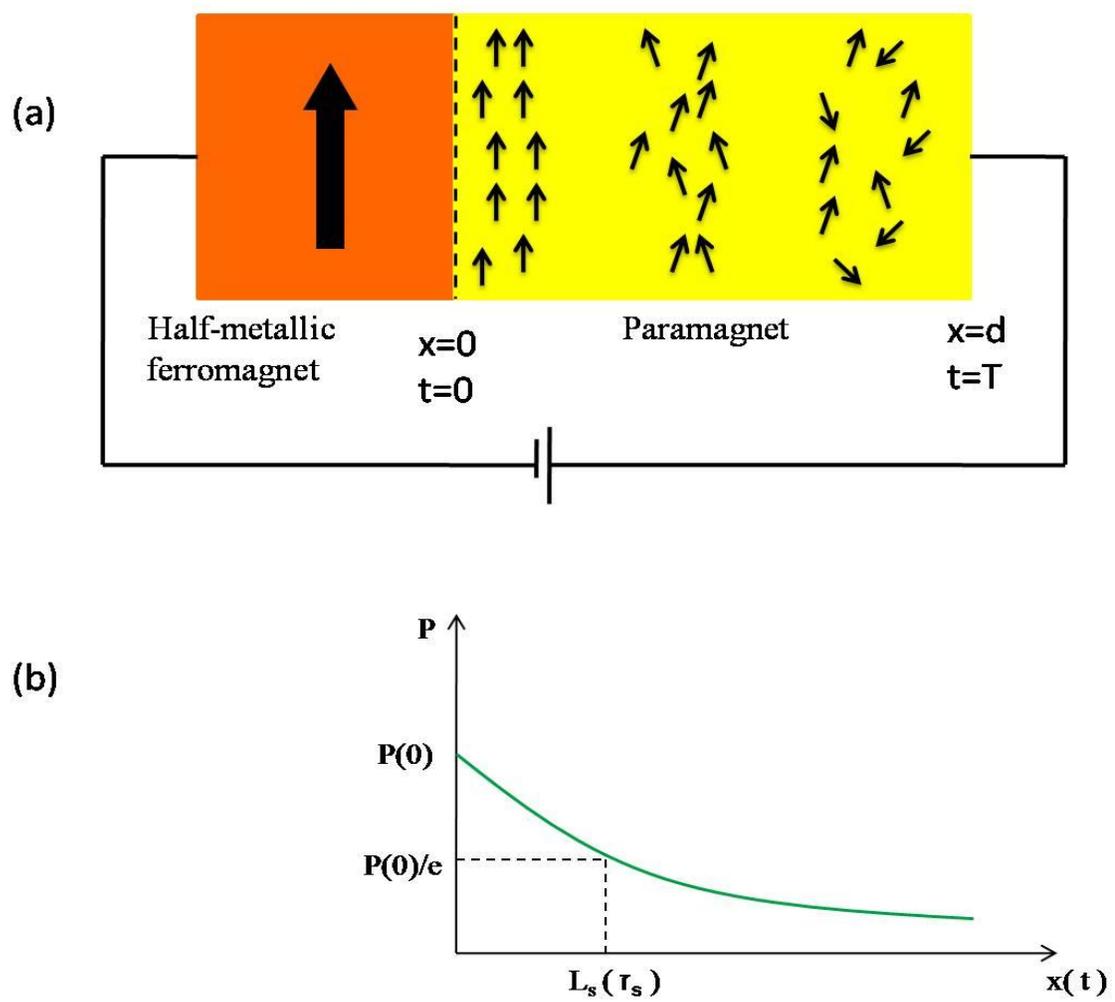

**Figure 2.** Illustration of electrical spin injection (a) and spin relaxation (b). The magnitude of the ensemble spin polarization decreases with time and distance implying spin relaxation.

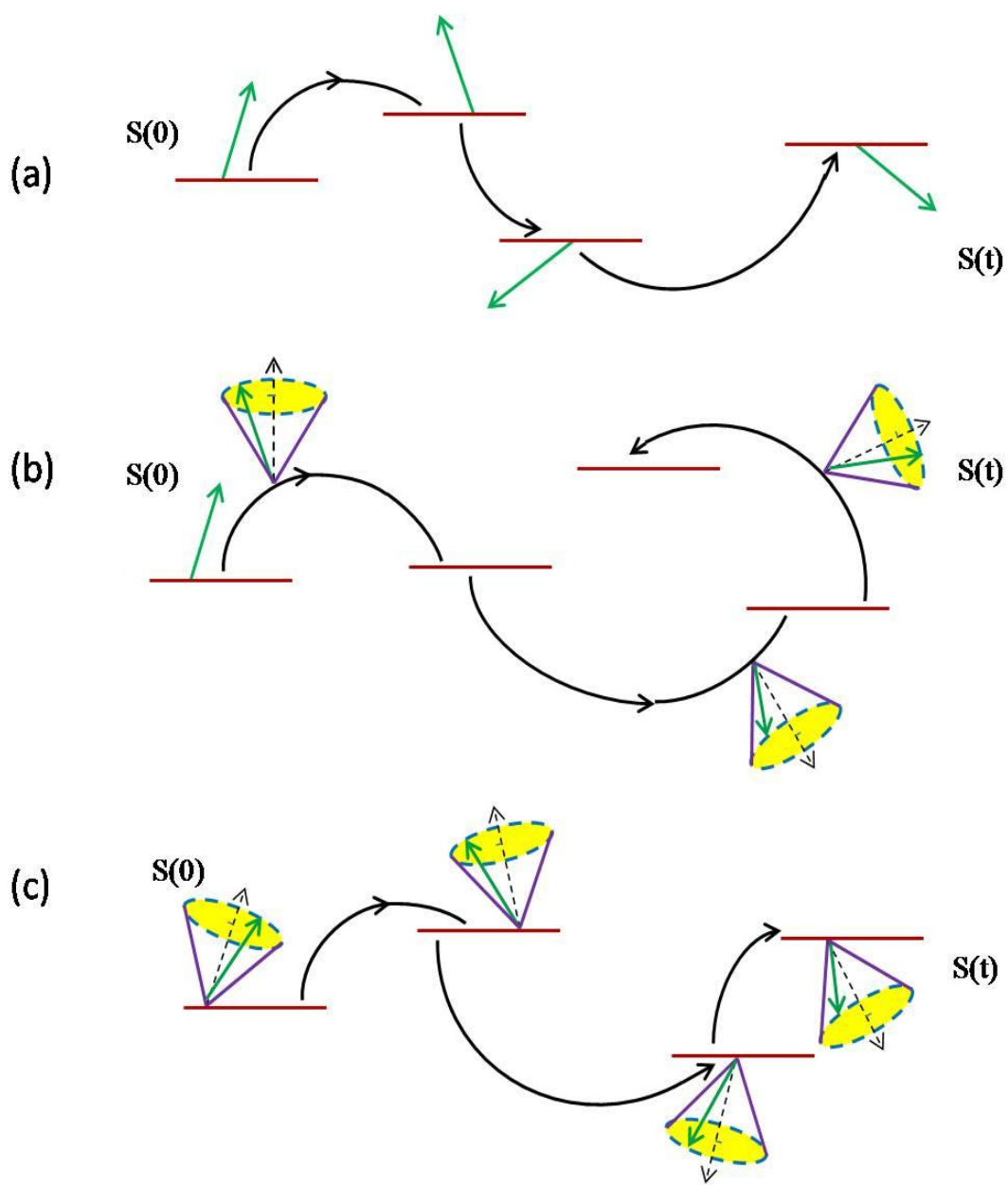

**Figure 3.** Schematic description of various spin relaxation mechanisms. Top, middle and the bottom panels represent spin relaxation via Elliott-Yafet, D'yakonov-Perel' and hyperfine interaction respectively.

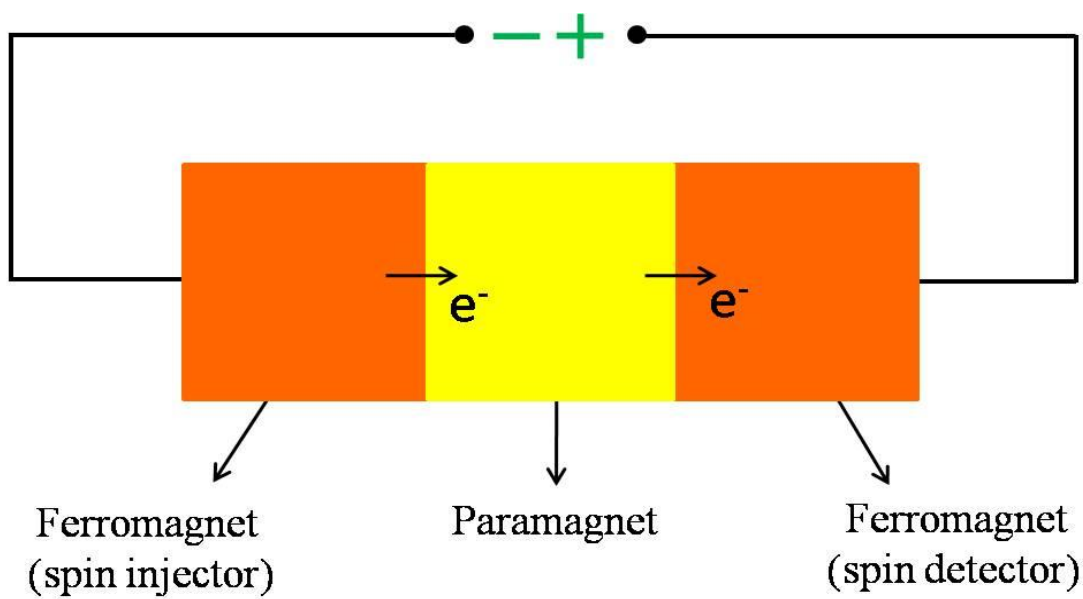

**Figure 4(a)** Schematic depiction of a spin valve. The spin valve is a trilayered structure where the paramagnetic spacer material is sandwiched between two ferromagnets of different coercivities. One of the ferromagnets acts as a spin injector and the other one act as the spin detector.

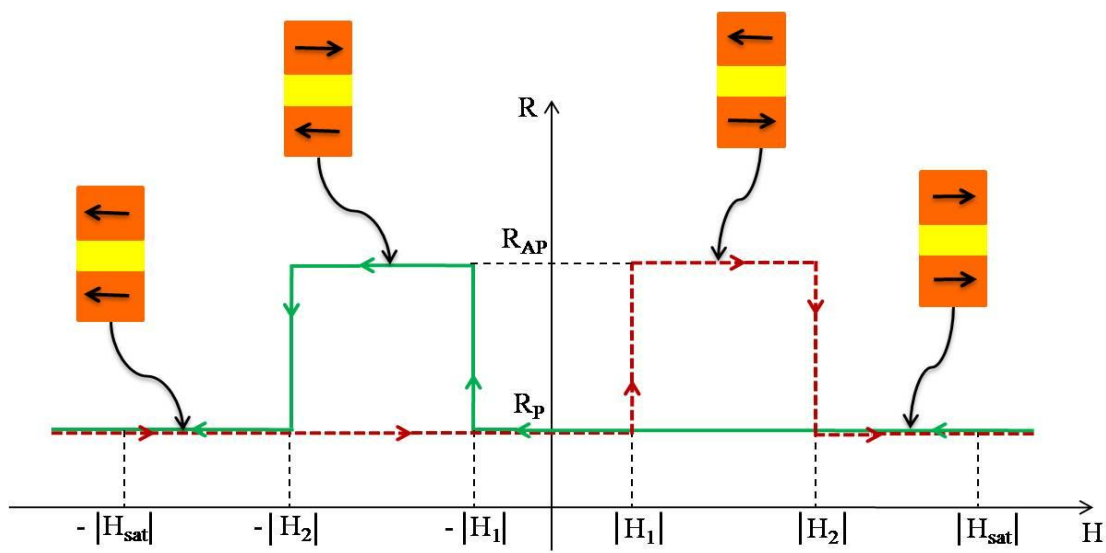

**Figure 4(b)** Pictorial representation of a spin valve response. The coercivities of the two ferromagnets are $|H_1|$ and $|H_2|$ as described in text. This is an example of the positive (or normal) spin valve effect since resistance corresponding to antiparallel magnetization is higher than that of parallel configuration.

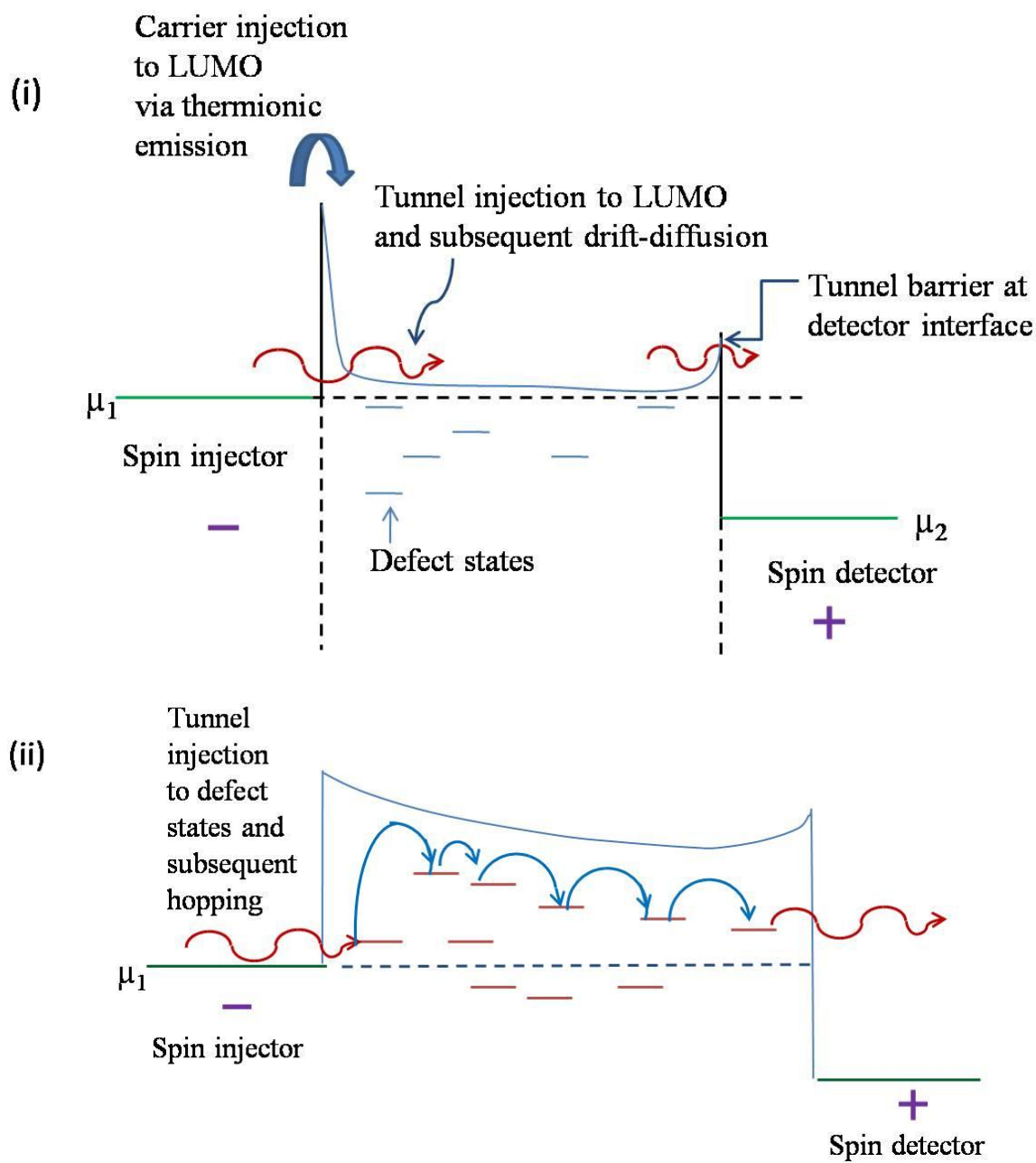

**Figure 4(c)** Physical model describing the modified Julliere formula. Tunnel or thermionic injection followed by drift-diffusion (top) and hopping transport (bottom). Spin polarization decreases exponentially with distance from the point of injection.

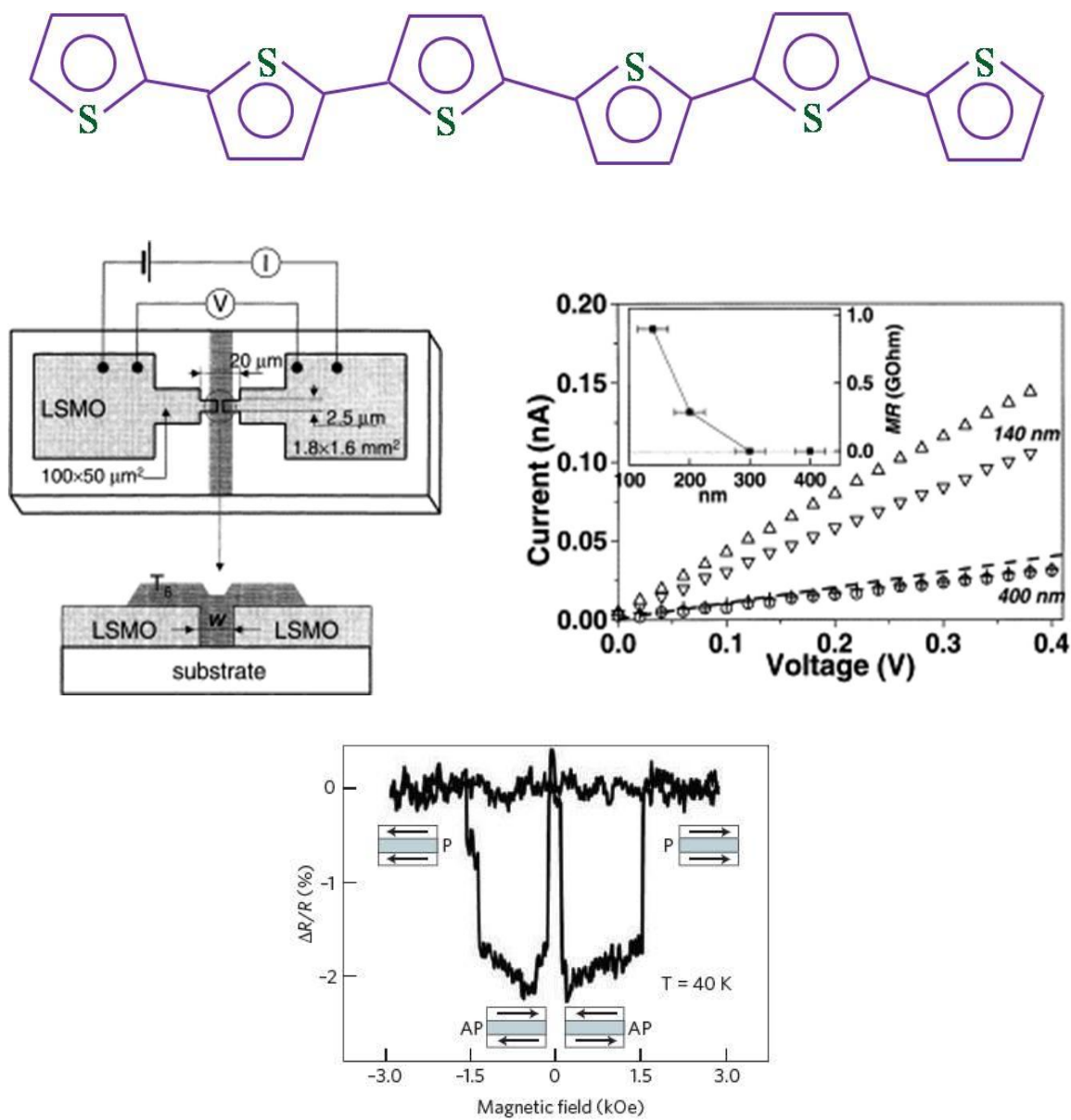

**Figure 5.** Spin valve experiment with sexithienyl. (*Top*) Molecular structure of $T_6$. Six linked thienyl units form a molecular axis. (*Middle left*) The device schematic and (*Middle right*) I–V characteristics as a function of magnetic field for 140 and 400 nm channel lengths. Down triangles and circles correspond to $H = 0$, while up triangles and crosses to $H = 3.4$ kOe. The dashed line indicates the expected slope for 400 nm junction as calculated from 140 nm junction by assuming a linear resistance increase versus channel length. The inset indicates MR as function of $w$, where MR $= R(0) - R(3.4$ kOe$)$. (after Dediu et al. [11]). (*Bottom*) Magnetoresistance loop, demonstrating inverse spin valve effect for a LSMO (20 nm)/ $T_6$ (100 nm)/$Al_2O_3$ (2 nm)/Co (20 nm) vertical spin-valve device measured at 40 K (after Dediu et al. [14])

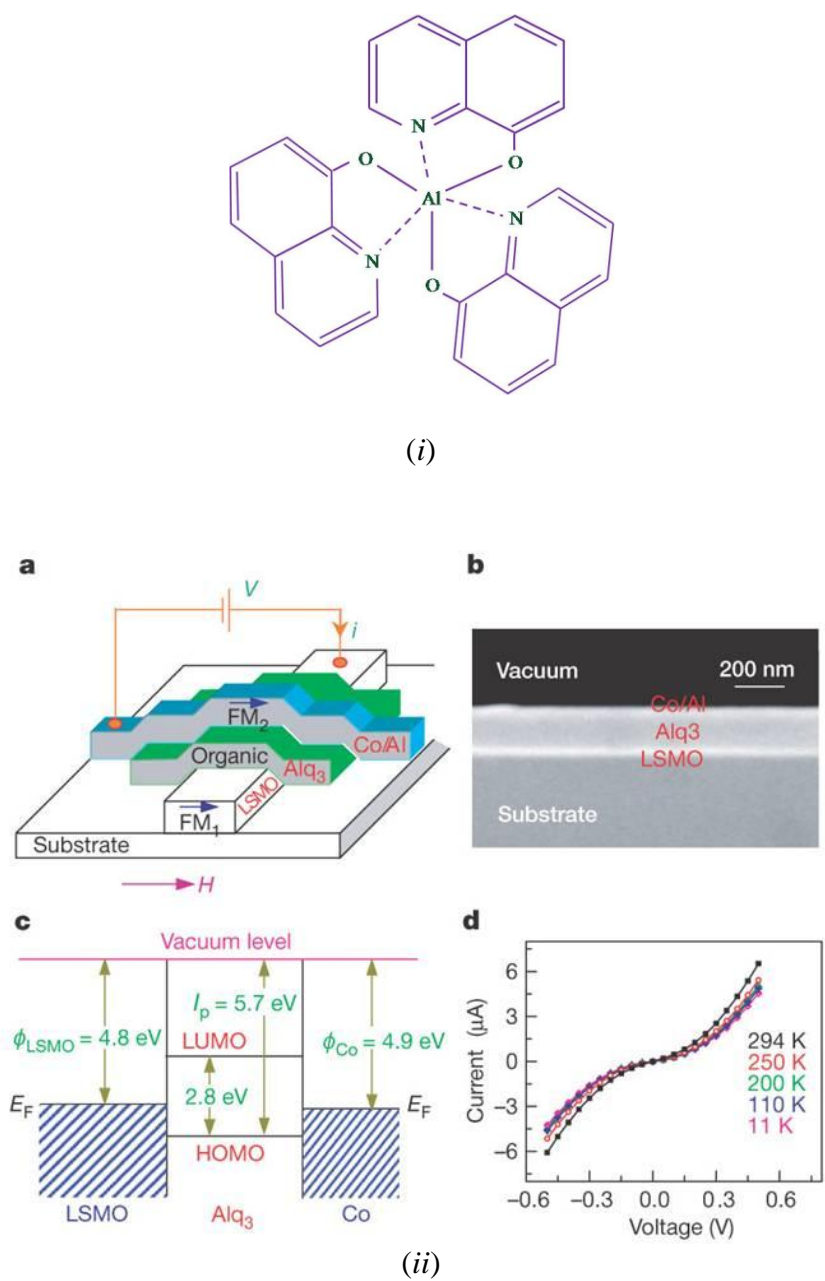

**Figure 6.** (*i*) Molecular structure of $Alq_3$. (*ii*) (a) The structure and I-V characteristics of the Ni- $Alq_3$-Co spin-valves. Spin-polarized current *i* flows from FM1 (LSMO), through the OSE spacer ($Alq_3$), to FM2 (Co) when a positive bias *V* is applied. An in-plane magnetic field, *H*, is swept to switch the magnetization directions of the two FM electrodes separately. (b) Scanning electron micrograph of a functional organic spin-valve. (c) Schematic band diagram of the OSE device in the rigid band approximation. (d) *I–V* response of the device with *d* = 200 nm at several temperatures. (after Xiong et al. [108]).

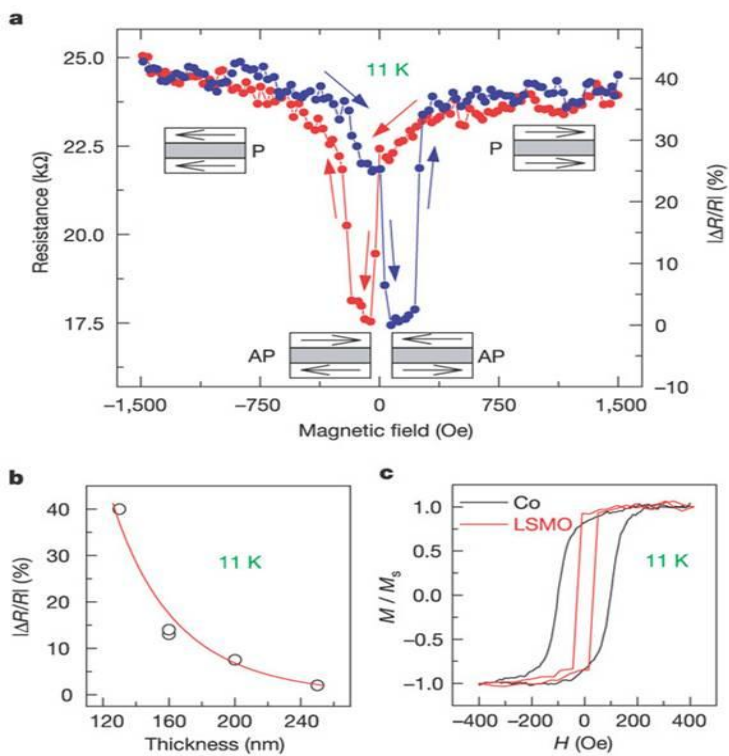

(*iii*)

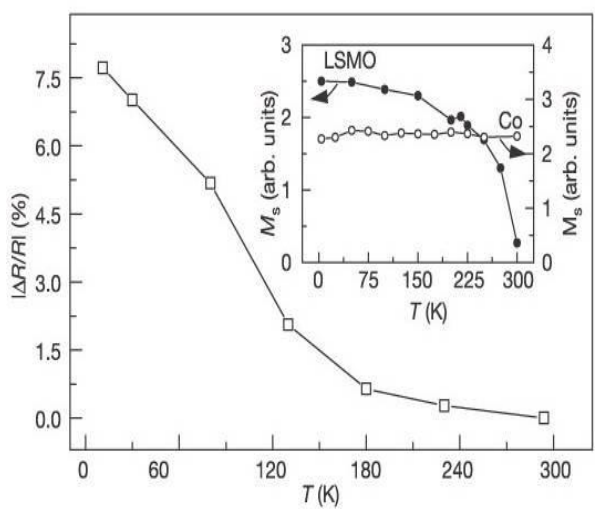

(*iv*)

**Figure 6 (contd.)** (*iii*) (a) Inverse spin valve response of LSMO/ $Alq_3$/Co at 11 K. (b) Spin valve signal as a function of $Alq_3$ thickness *d*. (c) Magnetic hysteresis loops measured using MOKE for the LSMO and Co electrodes (*iv*) $\Delta R/R$ measured at $V$ = 2.5 mV as a function of temperature. The inset shows the magnetizations of the Co and LSMO electrodes versus T, measured using MOKE. (after Xiong et al., [108]).

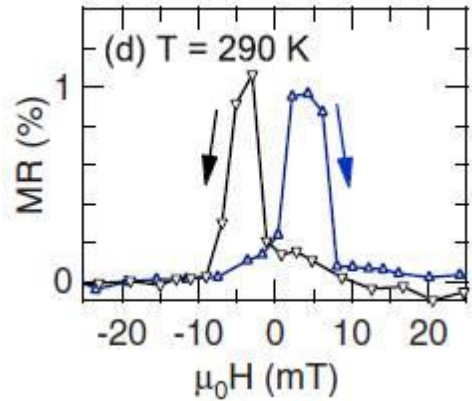

(*a*)

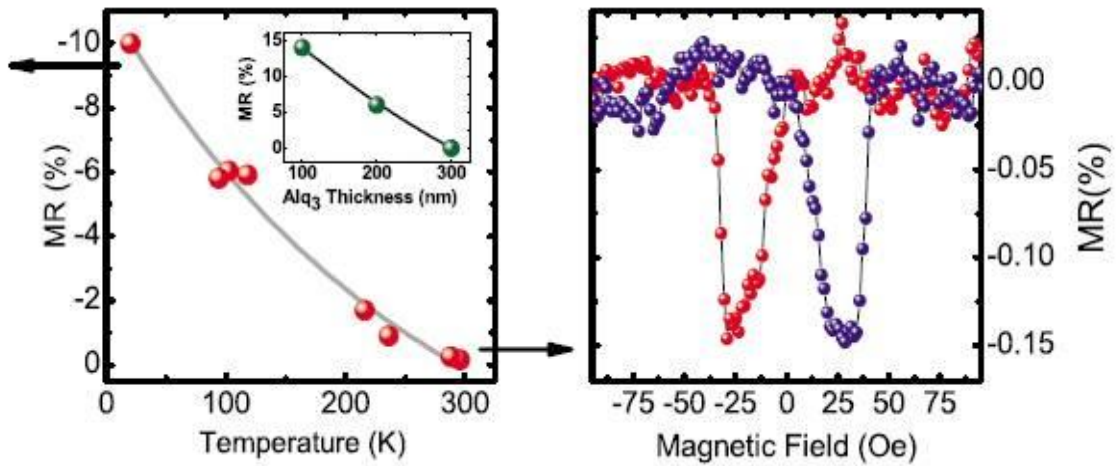

(*b*)

**Figure 7.** **(a)** Magnetoresistance data of Fe/$Alq_3$/Co device taken at 290 K. (after Liu et al., [117]). **(b)** (Left) Magnetoresistance values as a function of temperature ($LSMO/Alq_3/Al_2O_3/Co$). The MR decreases with increasing temperature but persists up to room temperature. (Right) Room-temperature inverse spin-valve effect in $LSMO/Alq_3/Al_2O_3/Co$. (after Dediu et al., [110]).

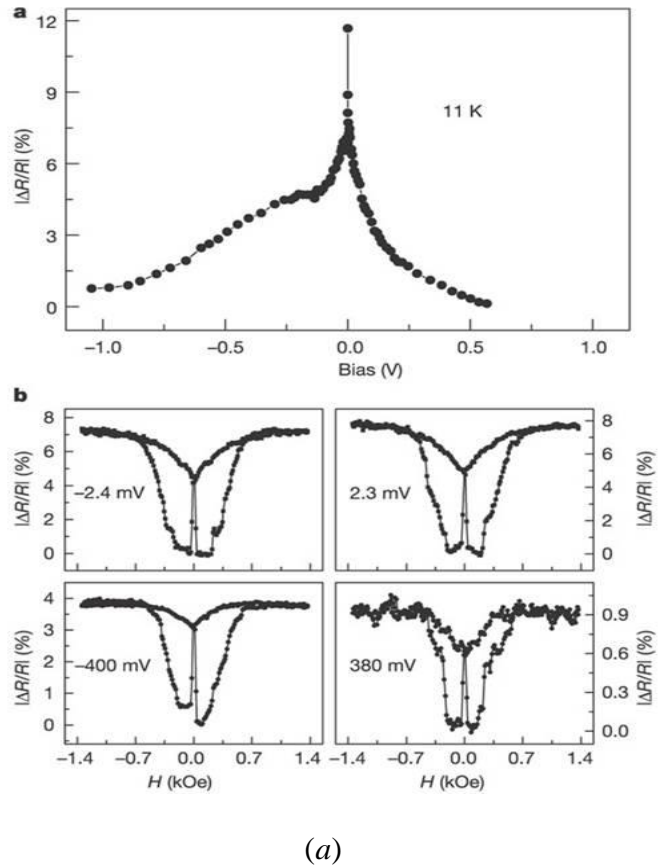

(*a*)

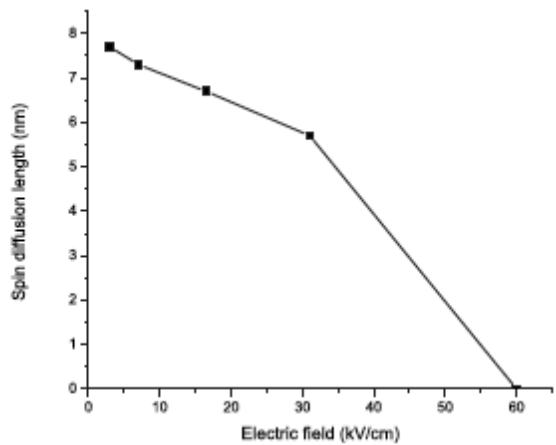

(*b*)

**Figure 8 (a)** $\Delta R/R$ for a LSMO/Alq3/Co device in Figure 6, measured at 11 K, as a function of *V*. (*b*) Spin-valve response at four different *V* values. (after Xiong et al., [108] **(b)** Spin diffusion length as a function of electric field for $Alq_3$ nanowires at a temperature of 1.9 K. (after Bandyopadhyay et al., [84]).

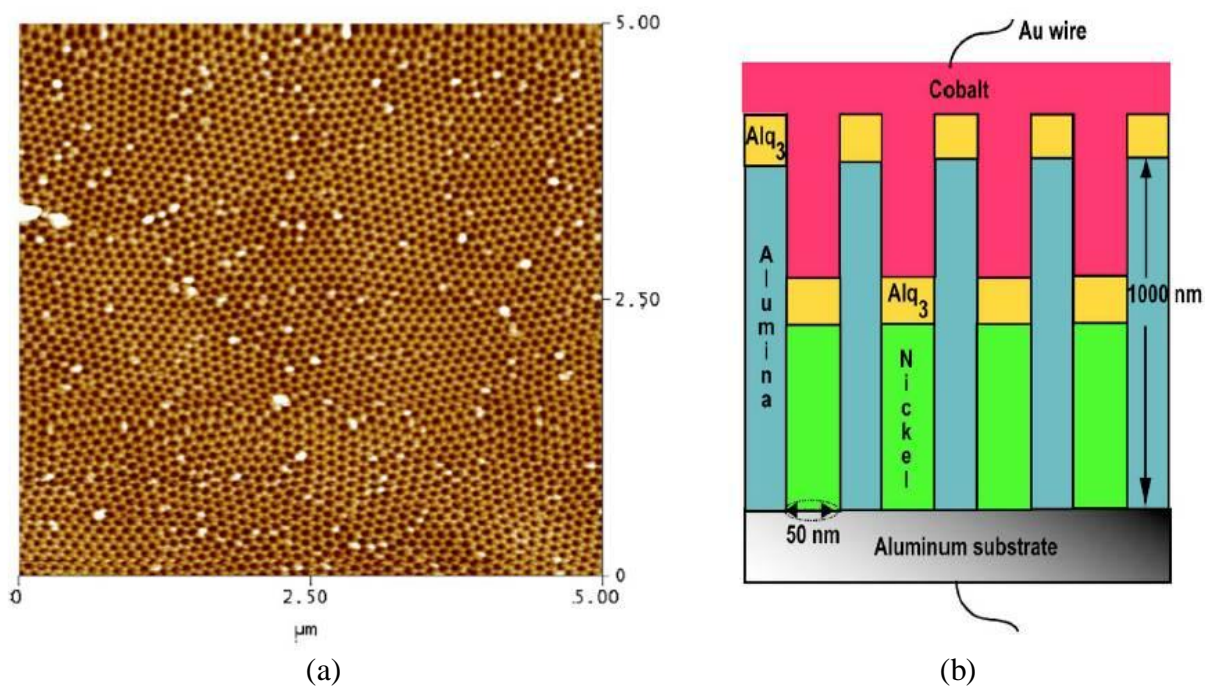

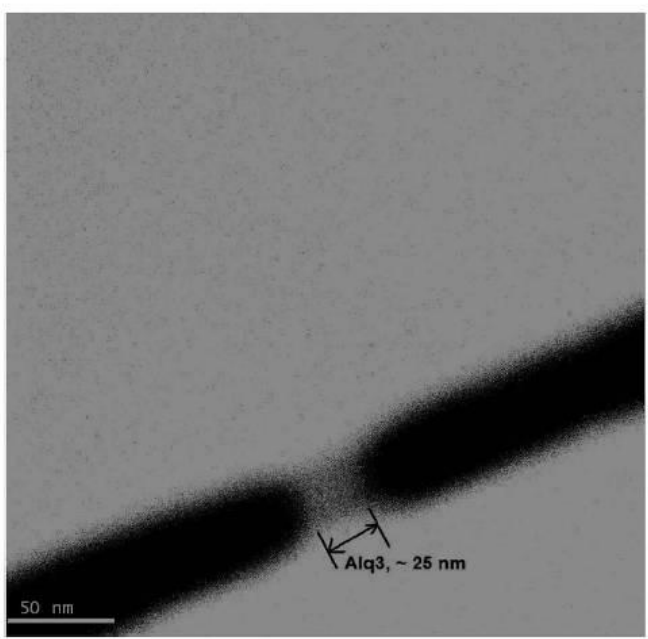

**Figure 9. (a)** Atomic force microscopy image of the top surface of the alumina template formed by anodization using 3% oxalic acid at 40 V dc. Pore diameter ~50 nm. **(b)** Schematic representation of the nanowire organic spin-valve device. The nanowires are hosted in an insulating porous alumina matrix and are electrically accessed from each end. The magnetic field is applied along the axis of the wire. **(c)** Transmission electron micrograph of a single nanowire showing the $Alq_3$ layer sandwiched between the cobalt and nickel electrodes. This image was produced by releasing the nanowires from the alumina host by dissolution of alumina in phosphoric acid and capturing the nanowires on TEM grids. (after Pramanik et al., [118]).

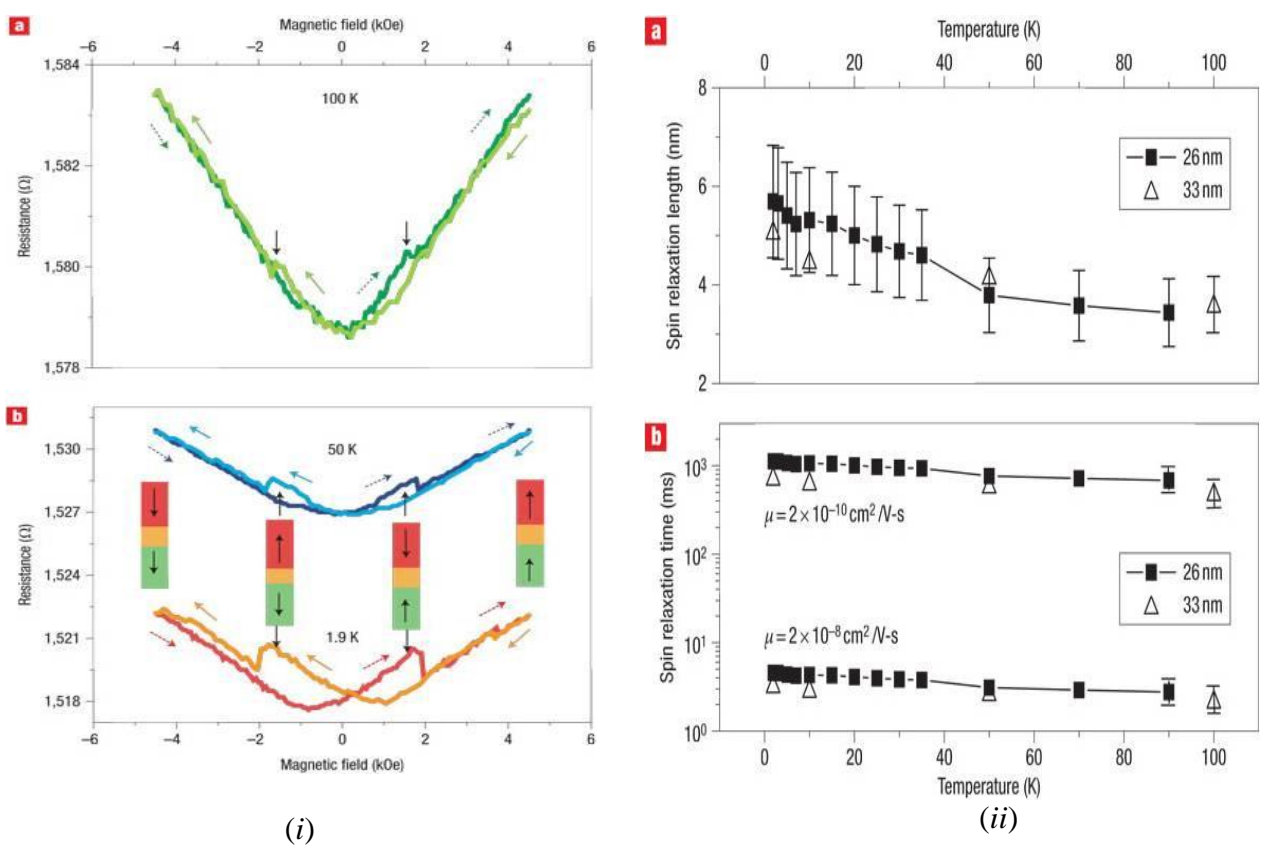

**Figure 10** (*i*) Magnetoresistance traces of Ni/$Alq_3$(33 $nm$)/Co nanowire spin valves at different temperatures. a,b, Traces with the magnetic field parallel to the axis of the wires at a temperature of 100 K (a) and at temperatures of 1.9 K and 50 K (b). The colored solid and broken arrows represent reverse and forward scans of the magnetic field, respectively. The parallel and antiparallel configurations of the ferromagnetic layers are shown within the corresponding magnetic field ranges in Fig. b. (*ii*) Temperature dependence of the spin relaxation length and time. a, The measured spin diffusion length as a function of temperature for two samples containing organic spacer layers of thicknesses 26 nm and 33 nm. The error bars accrue from the +/- 5nm uncertainty in the organic layer thickness in each case. b, The spin relaxation time as a function of temperature. The two curves correspond to the maximum and minimum values at any temperature. *The large difference between the maximum and minimum values is entirely due to the two orders of magnitude spread in the reported mobility of carriers in $Alq_3$*. (after Pramanik et al., [100]).

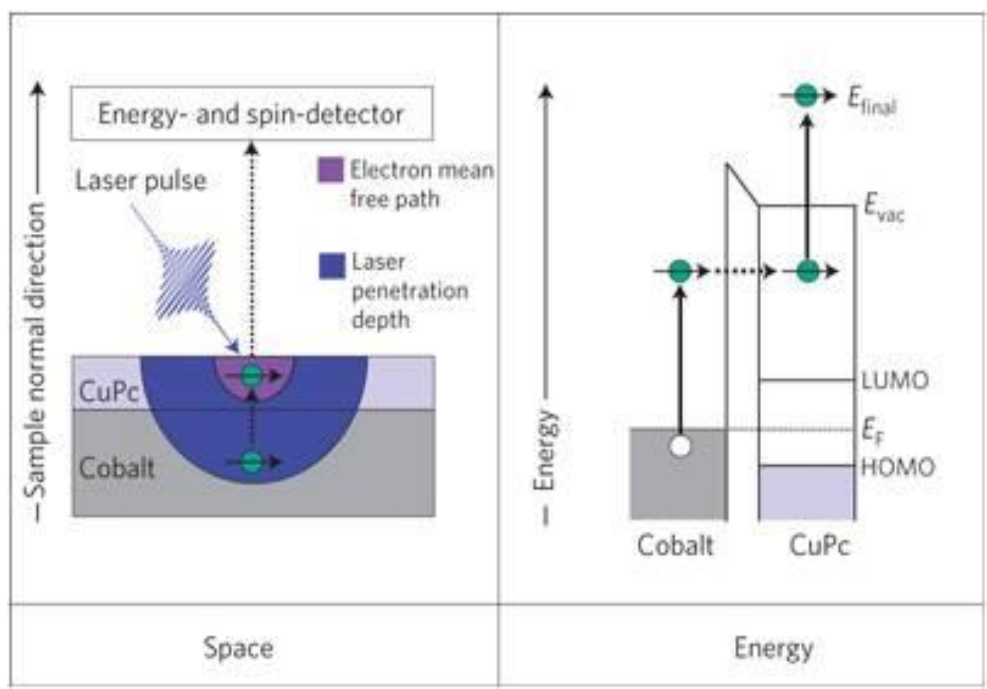

**Figure 11. Conceptual principle of the TPP experiments.** The sample, constituted of a cobalt thin film covered with a homogeneous CuPc film of variable thickness, is illuminated with pulsed laser light with photon energy 3.1 eV. The laser penetration depth (blue shaded area) is much larger (at least ten orders of magnitude) than the inelastic mean free path of the electrons excited by the laser (violet shaded area). As a consequence, a first photon generates spin-polarized electrons in the cobalt film, and only those spin-polarized electrons that reach the surface region in CuPc are subsequently photoemitted by absorbing a second photon from the laser pulse. The energy and the spin component along the cobalt magnetization direction of the photoemitted electrons are analysed. Energetically, the electrons are excited by the first pulse in intermediate states lying between the Fermi and the vacuum level of the heterojunction. A second photon gives to some of those excited electrons enough energy to be photoemitted (2PPE process). Before being photoemitted, the electrons must travel from cobalt into CuPc, or in other words they must be injected from cobalt into molecular orbitals of CuPc lying above the LUMO level. (after Cinchetti et al., [106]).

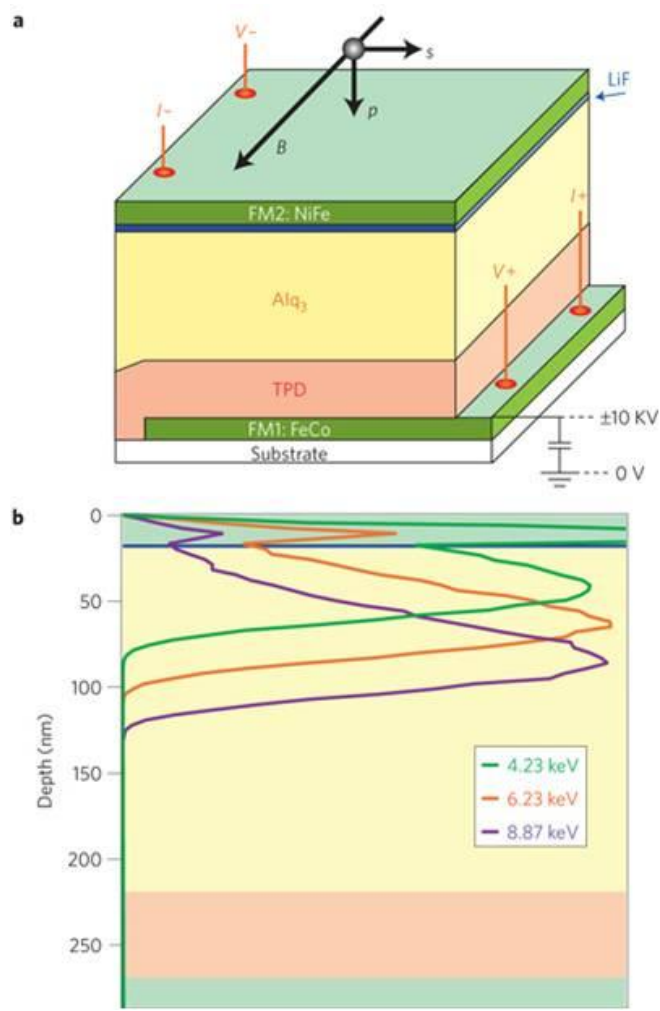

**Figure 12. Schematic diagram of the $\mu SR$ experiment.** (**a**) A schematic diagram showing the structure of the device. FM1 (FM2): ferromagnetic layer 1 (2). (**b**) Depth profile showing the calculated probability that a positively charged muon with an implantation energy of 4.23, 6.23 or 8.87 keV (green, orange or purple line) comes to rest at a certain depth within the device (so-called stopping or implantation profile). (After Drew et al., [105])